\documentclass[twocolumn,
]
{revtex4-2}

\usepackage{graphicx}
\usepackage{dcolumn}
\usepackage{bm}
\usepackage{amsmath}
\usepackage{amssymb}
\usepackage{hyperref}
\usepackage{color}
\usepackage{verbatim}
\graphicspath{{fig}}
\usepackage{overpic}
\usepackage{subfigure}
\usepackage{bm}

\newcommand{\BB}{\mathbf{B}}

\newcommand{\MM}{\boldsymbol{\mathsf{M}}}
\newcommand{\xx}{\mathbf{x}}
\newcommand{\vv}{\boldsymbol{v}}
\newcommand{\uu}{\boldsymbol{u}}
\newcommand{\aaa}{\boldsymbol{a}}
\newcommand{\bb}{\boldsymbol{b}}
\newcommand{\xxo}{\mathbf{x}_0}
\newcommand{\TT}{\boldsymbol{\mathsf{T}}}
\newcommand{\biso}{\hat{\mathbf{b}}_{\rm ext}}
\newcommand{\XX}{\mathbf{X}}
\newcommand{\YY}{\mathbf{Y}}
\newcommand{\gal}{g_\alpha}
\newcommand{\gbe}{g_\beta}
\newcommand{\gp}[1]{g_{\phi_{#1}}}
\newcommand{\VV}{\boldsymbol{V}}

\makeatletter
\def\@email#1#2{%
 \endgroup
 \patchcmd{\titleblock@produce}
  {\frontmatter@RRAPformat}
  {\frontmatter@RRAPformat{\produce@RRAP{*#1\href{mailto:#2}{#2}}}\frontmatter@RRAPformat}
  {}{}
}%
\makeatother

\begin{document}
\title{A topological approach to magnetic nulls}

\author{B. Y. Israeli} 
\affiliation{Weizmann Institute of Science\\
Department of Physics of Complex Systems\\
Rehovot 7610001 Israel}

\author{C. B. Smiet} 
\affiliation{Ecole Polytechnique Fédérale de Lausanne (EPFL) \\ 
Swiss Plasma Center (SPC) 
\\ CH-1015 Lausanne, Switzerland
}

\begin{abstract}
  Magnetic nulls are locations where the magnetic field vanishes.
  They are the location of magnetic reconnection, and they determine to a large degree the magnetic connectivity in a system. 
  We describe a novel approach to understanding movement, appearance, and disappearance of nulls in magnetic fields.
  This approach is based on the concept of \emph{isotropes}, or lines where the field direction is constant. 
  These lines are streamlines of a vector field whose flux is sourced by the topological indices of nulls, and can be conceptualized as corresponding ``lines of force'' between nulls. 
  We show how this topological approach can be used to generate analytical expressions for the location of
  nulls in the presence of external fields for dipoles and for a field defined by the Hopf fibration. 
\end{abstract}

\maketitle

\section{introduction}
In Sherlock Holmes' `curious incident of the dog in the night-time'~\citep{doyle1979adventure}, it is the absence of something (the bark of a dog) which allows for the reconstruction of the bigger picture, and the solving of the case.
In astrophysical plasmas it is similar absences of magnetic field, vanishing in topologically protected magnetic nulls, that give insight concerning  overall connectivity, and ultimately the global plasma dynamics.
In the solar corona, the location of magnetic nulls impacts energy propagation~\citep{candelaresi2016effects}, and magnetic nulls form the loci of magnetic reconnection~\citep{priest1996magnetic, schindler1988general, greene1988geometrical, lau1990three}.
The interaction of the solar wind with planetary magnetic fields gives rise to magnetic nulls~\citep{cowley1973qualitative}.
On Earth, magnetic confinement fusion devices such as the FRC~\citep{tuszewski1988field} create a plasma-confining magnetic field containing two magnetic nulls.

The field around a magnetic null has a universal structure~\citep{parnell1996structure}
which has hitherto been elucidated by looking at the linearized field given by the matrix of partial derivatives (the Jacobian) $\mathsf{M}_{ij}=\partial_jB_i$ evaluated at the null location; $\xxo$ \citep{parnell1996structure, lau1990three, greene1988geometrical}.
By virtue of $\MM$ being the linearization of the magnetic field ($\MM\delta\xx=\BB(\xxo+\delta\xx)$), the eigenvectors of $\MM$ correspond to directions in which field lines asymptotically approach or leave the null \citep{parnell1996structure}.
The trace of $\MM$ (and therefore the sum of its eigenvalues) is zero by the solenoidal condition ($\nabla\cdot\mathbf{B}=0$).
In three dimensions, the eigenvectors therefore consist of one singular eigenvector, and two coupled eigenvectors whose eigenvalues are either complex conjugate or real, and are opposite in sign to the singular eigenvector.
The field line in the direction of the singular eigenvector is called the `spine' of the null, whereas the two coupled eigenvectors span a plane called the `fan plane'.
This universal structure is demonstrated in figure~\ref{fig:threenulls} (a) for a linear null with the spine along the $z$ axis and the fan in the $(x,y)$-plane.
We recommend~\citet{parnell1996structure} for a comprehensive overview of the above.

Aside from the above analysis in terms of linearized matrices, it has been recognized that nulls are topological in nature, and are singular points of the magnetic field~\citep{greene1988geometrical, greene1992locating}.
As such they have an associated topological (or Poincar\'e) index, and cannot disappear or appear except by bifurcations in which two or more opposite index points appear or disappear together.
This index is employed in locating magnetic nulls in numerical simulation~\citep{greene1992locating,olshevsky2020comparison} and from satellite cluster data~\citep{fu2020methods}.

In this paper we expand upon the foundational understanding of magnetic nulls laid down by~\citet{greene1988geometrical, greene1992locating},~\citet{lau1990three}, and~\citet{parnell1996structure, murphy2015appearance}, and introduce a method for finding and analyzing magnetic nulls that utilizes their topological nature without requiring information of the field in a large area. 
We introduce the concept of an isotrope field (from the Greek \emph{iso-}, 'same' and \emph{tropos}, `turn', `direction', or `way'), the one-form field that points in the direction in which the magnetic field direction remains unchanged.
We show how this simple-to-calculate field yields an intuitive global picture of the location and movement of magnetic nulls in complex fields.
Specifically we will examine the location of magnetic nulls in two toy models of astrophysical importance consisting of the combination of a localized magnetic field and an externally sourced constant field.
We derive the analytical expressions for the locations of the nulls, and visualize the merger of two nulls.
We further show that pairs of opposite index nulls interact as an external field is varied and may only form/annihilate on constrained surfaces.

While these calculations were performed without conscious analogy to other systems, these methods have precise analogs in other areas of physics, notably the study of condensed matter systems such as liquid crystals, ferromagnets, and superfluids, in which topological defects are of significant interest, and in topological field theory.
Numerous works in these fields utilize similar mathematical machinery for calculating topological indices~\citep{flandersVIApplicationsEuclidean1963,kurikDefectsLiquidCrystals1988,alexander2012nematic}, and some even make note of the utility of a vector field in expressing such calculations~\citep{blahaQuantizationRulesPoint1976,trebinTopologyNonuniformMedia1982,devegaClosedVorticesHopf1978}.
However, as the fields in these systems take values on compact manifolds (the sphere for three-dimensional ferromagnets for example), the link between topological index and paths of constant field direction is not emphasized.
Due to the magnetic field taking values in a vector space, and its resulting additive properties, the isotrope field and \emph{its integral curves} take on additional significance in describing evolving magnetic field structures, as will be discussed.

\section{The topological index of a null}
The most general topological method for classifying nulls is via the \emph{topological index} that is defined as follows: 
Let the \emph{director map} $g:\mathbb{R}^3\rightarrow S^2$ be the map that sends points in space to unit vectors in the direction of the field $\mathbf{B}$ (i.e. points on the unit sphere):
\begin{equation}\label{eq:director}
g(\mathbf{x}) = \frac{\mathbf{B}(\mathbf{x})}{|\mathbf{B}(\mathbf{x})|}.
\end{equation}
Let $D^{3}_{\xxo}$ be a ball centered on a null located at $\xxo$ and enclosing only that null. 
Restricting $g$ to the boundary $\partial D^3_{\xxo}$, we have a map from $S^2\rightarrow S^2$. 
The index of the null is then defined by:
\begin{equation}\label{eq:index1}
\mathrm{Ind}(\xxo)=\mathrm{Deg}(g|_{\partial D^3_{\xxo}}).
\end{equation}

\begin{figure}
    \begin{overpic}[width=\linewidth]{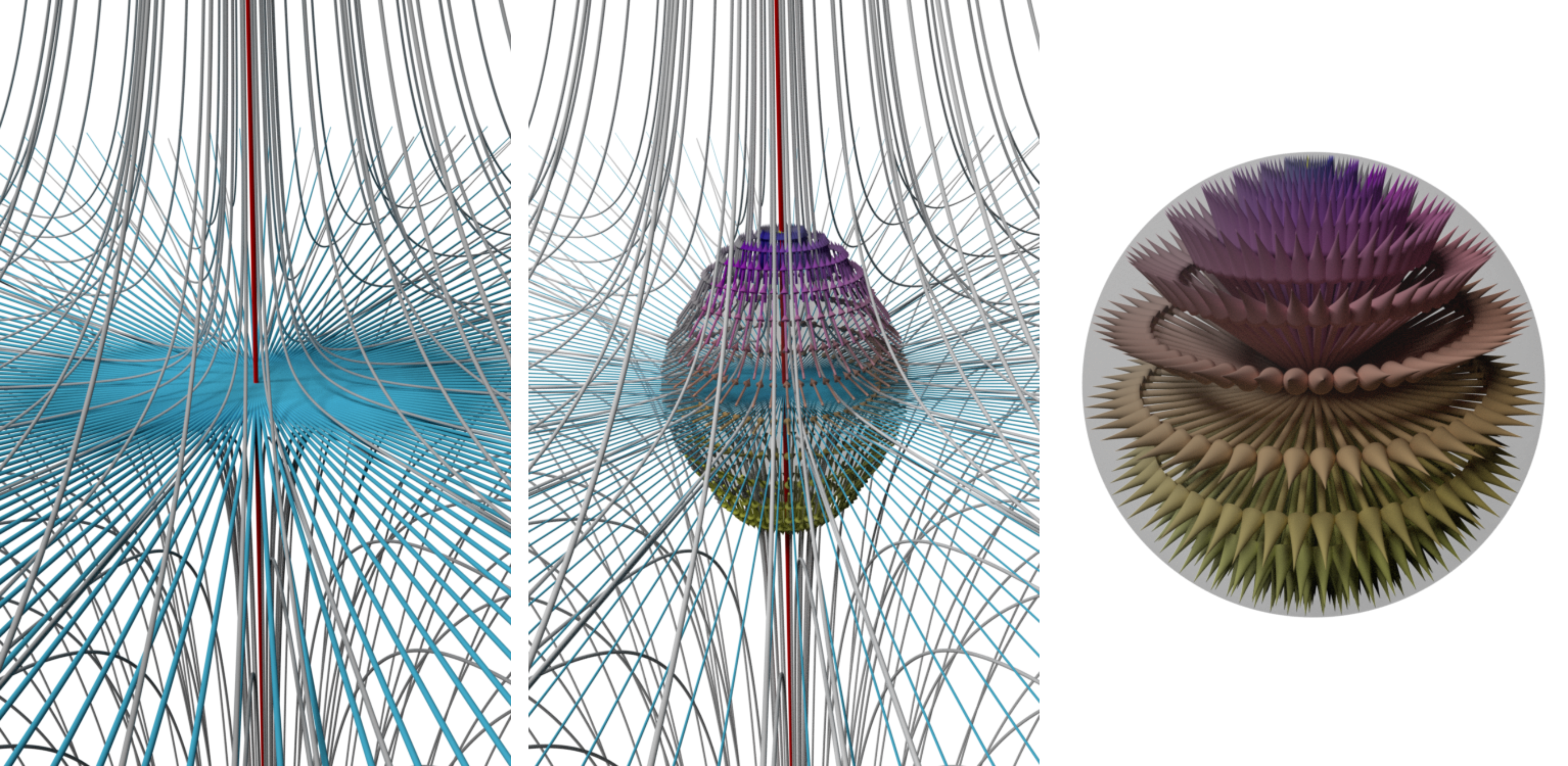}
    \put(1,45){\fcolorbox{white}{white}{\parbox{7pt}{\centering \textbf{a}}}}
    \put(34,45){\fcolorbox{white}{white}{\parbox{7pt}{\centering \textbf{b}}}}
    \put(67,45){\fcolorbox{white}{white}{\parbox{7pt}{\centering \textbf{c}}}}
    \end{overpic}
  \caption{
  The structure of a magnetic null and an illustration of the calculation of the topological index. (a) Magnetic field lines in the null configuration, with field lines of the spine in red, those of the fan plane in blue, and several nearby field lines in white.(b) Vectors of the magnetic field on a small sphere $\partial D^3$ (transparent, gray) colored from purple (north pole) to yellow (south pole). (c) The vectors of (a) translated to the origin and given unit length, to illustrate the mapping. 
  }\label{fig:threenulls}
\end{figure}

The degree of a map is integer-valued and can be interpreted as the (signed) number of times the sphere is covered. 
This is the three-dimensional generalization of the winding number that can be used to classify X-points and O-points in Poincar\'e sections~\citep{smiet2019mapping}. 
The concept of the degree of a mapping is visualized in figure \ref{fig:threenulls}, where (a) shows the magnetic field lines of a current-free null, with the spine (fan plane) in red (blue). Figure~\ref{fig:threenulls} (b) includes the sphere $\partial D^3_{\xxo}$ in transparent gray, and the vector field evaluated on its surface. The vectors range in color from from purple (north pole) to yellow (south pole). In (c) these vectors are mapped to the unit sphere: the north pole (purple arrows) is mapped to the north pole of $S^2$, and the south pole (yellow arrows) are mapped to the south pole. 
This map covers the range exactly once 'right side out', and therefore has degree +1 and index +1. 

The degree of a map is robust to perturbations of the region $D^3$, and can be calculated for the surface of an arbitrarily-shaped region $U\in\mathbb{R}^3$. 
If $U$ contains more than one isolated null, then the following theorem holds:
\begin{equation}\label{eq:indextheorem}
\mathrm{Deg}(g|_{\partial U})= \sum_{\xxo\in U} \mathrm{Ind}(\xxo).
\end{equation}
The degree of the map is the sum of the topological indices of the enclosed nulls.

Several other classification methods have been described for nulls, based on the properties of the matrix $\MM$ evaluated at the null. 
\citet{greene1992locating, greene1988geometrical} defined a positive (negative) null as one for which the determinant $|\MM|$ is positive (negative). 
Recall that the geometric meaning of the determinant is the signed volume scaling factor of the linear transformation produced by $\MM$.
A negative determinant thus maps the surrounding area 'inside out', and corresponds to a mapping with degree -1. This definition is consistent with the topological index classification. 
Another definition, used in~\citet{olshevsky2020comparison}, is the sign of the product of eigenvalues of $\MM$, which is trivially equivalent to Greene's and our definition since that equals the determinant. 
The other classifications rely on the signs of the eigenvalues of $\MM$. 
These must sum to zero, and since $\MM$ is real-valued, there is one singular (real) eigenvalue, and two coupled eigenvalues with real part of opposite sign to the singular eigenvalue.
\citet{lau1990three} classified nulls by the sign of the singular eigenvector (corresponding to the spine), with Type A nulls those where the singular eigenvalue is positive, and type B nulls where it is negative.
The determinant, equalling the product of eigenvalues, has the sign of the singular eigenvalue. Type A nulls therefore have index +1, and type B have index -1.
~\citet{parnell1996structure} used the sign of the real part of the two coupled eigenvectors to classify nulls thus labeling a null where the field is directed outwards on the fan plane as positive, though it has index -1 and would be classified as negative by~\citet{greene1988geometrical, olshevsky2020comparison}.

\section{Isotrope Lines}
We can evaluate the degree of the mapping, by counting covers of $S^2$ on the boundary of a region $U\subset\mathbb{R}^3$: 
\begin{equation}\label{eq:intdeg}
      \mathrm{Deg}(g|_{\partial U}) = \sum_{\xx_i\in U}\mathrm{Ind}(\xx_i) = \frac{1}{4\pi}\int_{g_*\partial U}\omega
\end{equation}
where we have defined some properly normalized area two-form $\omega$ on $S^2$, and where $g_*\partial U$ is the push forward with respect to $g$ of the boundary of $U$.
Note that the choice of $\omega$ may be made arbitrarily to suit the problem at hand as long as it is normalized to $\int_{S^2}\omega=4\pi$.

To transfer our notion of area on $S^2$ (defined by $\omega$) to an object that can be integrated over a surface in $\mathbb{R}^3$, we take the pull-back through the map $g:\mathbb{R}^3\rightarrow S^2$ of the area two-form $\omega$:
\begin{equation}
    \mathrm{Deg}(g|_{\partial U})=\frac{1}{4\pi}\int_{\partial U}g^*\omega.
\end{equation}
Taking the Hodge dual of the pull-back of this two-form on $\mathbb{R}^3$ yields a one-form (a covector field):
\begin{equation}
    \vv=\star g^*\omega.
\end{equation}
This can be integrated over the surface $\partial U$: 
\begin{equation}
    4\pi\sum_{\xx_i\in U}\mathrm{Ind}(\xx_i)=\int_{\partial U}\vv\cdot da.
\end{equation} 

The vector field $\vv$, being dual to the pull-back of the area two-form on $S^2$, has a very special and interesting interpretation: the two-form encodes the directions in which the angle of the magnetic field changes, and $\vv$ \emph{is the vector along which the field direction remains constant} (as proved below).
As equation~\eqref{eq:director} constitutes a map from a 3 dimensional manifold to a 2 dimensional manifold, this map loses a degree of freedom, and at any point in $\mathbb{R}^3$ (except at nulls) there is a direction in space in which the direction of the magnetic field remains constant.
We call the resulting paths of constant direction the \emph{isotropes} (from the Greek \emph{iso-}, 'same' and \emph{tropos}, `turn', `direction', or `way') of the magnetic field.

Isotropes cannot start or end except at nulls, where an isotrope of each direction on $S^2$ must start or end.
Similar to the electric field of a point charge, the field $\vv$ encodes the topological charge. 
Each isotrope corresponds with a given infinitesimal area (the two-form that is its dual) of the target $S^2$ carrying local information of the global index calculation.

That $\vv$ points in the direction where the angle of the magnetic field remains constant can be seen as follows. Working in arbitrary coordinates $(\alpha,\beta)$ on $S^2$ with $\omega=d\alpha\wedge d\beta D(\alpha,\beta)$ and $g(\xx)=(\gal,\gbe)$, we can derive the general form of $\vv$ and verify it indeed lies tangent to the isotrope lines.
\begin{multline}
    \int_{g_*\partial U}\omega 
    =\int_{g_*\partial U}d\alpha d\beta D(\alpha,\beta)\\
    =\int_{\partial U}dxdy
    \begin{vmatrix}
      \partial_x\gal & \partial_x \gbe \\
      \partial_y\gal & \partial_y \gbe
    \end{vmatrix}
    D(\gal,\gbe)
    \equiv\int_{\partial U}\vv\cdot da.
\end{multline}
Using the relation of the Hodge dual to the vector triple product in $\mathbb{R}^3$, for any two vectors $\aaa,\bb$:
\begin{multline}
    \vv\cdot(\aaa\times \bb)
    =g^*\omega(\aaa,\bb)
    =\begin{vmatrix}
      \partial_{\aaa} \gal(\xx) & \partial_{\aaa} \gbe(\xx) \\
      \partial_{\bb} \gal(\xx) & \partial_{\bb} \gbe(\xx)
    \end{vmatrix}
    D(\gal(\xx),\gbe(\xx))\\ 
    =(\nabla\gal\times\nabla\gbe)\cdot(\aaa\times \bb)D(\gal,\gbe)
\end{multline}
So $\vv$ can be written as:
\begin{equation}
\label{eq:isotropefield}
    \vv=(\nabla\gal\times\nabla\gbe)D(\gal,\gbe).
\end{equation}
We can immediately see that $g:\xx\mapsto (\gal,\gbe)$ is constant on streamlines of $\vv$, so the streamlines of $\vv$ are by definition the previously defined isotropes of $\mathbf{B}$.

While we have not encountered equation \ref{eq:isotropefield} in the literature, another form can be found in various works~\citep{blahaQuantizationRulesPoint1976, trebinTopologyNonuniformMedia1982, devegaClosedVorticesHopf1978}, which has the advantage of simplicity while obscuring the relation to the pullback:
\begin{equation}
\label{eq:litisotrope}
    \vv=\frac{1}{2} \varepsilon_{i j k} \hat{b}_i\left(\nabla \hat{b}_j \times \nabla \hat{b}_k\right).
\end{equation}
We demonstrate the equivalence of these expressions in Appendix \ref{app:lit}.

Specifically for three dimensions, we will now choose the standard coordinates $(\theta, \phi)$ on $S^2$ with  $\omega=d\theta\wedge d\phi\sin\phi$, and write $g(\mathbf{x}) = \left(g_\theta(\mathbf{x}), g_\phi(\mathbf{x})\right)$ where
\begin{equation}
\begin{split}\label{eq:gfns}
    g_\theta(\mathbf{x})=&\arctan\left(B_y(\mathbf{x})/B_x(\mathbf{x})\right)\\
    g_\phi(\mathbf{x})=&\arctan\left(\sqrt{B_x^2(\xx) +B_y^2(\xx)}/B_z(\xx)\right).
\end{split}
\end{equation}
The equation for the isotrope field $\vv$ then becomes: 
\begin{equation}\label{eq:isotropes}
    \vv=(\nabla g_\theta\times\nabla g_\phi)\sin g_\phi.
\end{equation}

The field $\vv$ is easily evaluated, and can be integrated to find nulls in vector fields. 

A similar derivation may be performed in any number of dimensions and will produce a field with equivalent properties, for which a novel general expression is derived in Appendix \ref{app:ndim}. In particular, in two dimensions and taking an angular coordinate $\theta$ on $S^1$, the isotrope field is

\begin{equation}
    \vv=
    \begin{pmatrix}
        0 & 1\\
        -1 & 0
    \end{pmatrix}
    \cdot \nabla g_\theta.
\end{equation}

\subsection{Movement of nulls in a changing field}
The concept of isotropes also gives an alternative intuitive derivation of the movement of a null as the field changes. 
It was derived by~\citet{murphy2015appearance} that the velocity of a null is given by the equation: 
\begin{equation}
  \label{equ:velocity}
  \uu = - \MM^{-1} \cdot \frac{\partial \BB}{\partial t}.
\end{equation}

Let there be an isolated null at $\mathbf{x}_0$. 
Due to some process (plasma dynamics, external coil being switched on, etc), the magnetic field $\BB=\BB_0$ changes by $\delta \BB$. 
The field is no longer zero at $\mathbf{x}_0$, but the null cannot have disappeared. 
Recall that a null must have isotropes of every direction converge on it, including an isotrope in the direction opposite to $\delta \BB$. 
Along that isotrope of $\BB_0$, there will be a point where $\BB_0$ and $\delta \BB$ cancel, and that is where the null has moved to. 

Mathematically we can express this as follows: we need to find the $\delta\xx$ where the magnetic field is opposite to $\delta \BB$. 
Since $\MM$ encodes the linearized field at the null, $\MM\cdot \delta\xx \simeq \mathbf{B}(\xxo +\delta\xx)$, we can find the position of cancellation (and the isotrope corresponding to $-\delta\BB$) by inverting $\MM$:
\begin{equation}
  \delta\xx \simeq -\MM^{-1}\cdot \delta\BB.
\end{equation}
The change in the magnetic field is given by $\delta \BB = \int_{t=0}^{\delta t} \frac{\partial\BB}{\partial t} dt$, and the change in position by $\delta\xx = \int_{t=0}^{\delta t} \uu dt$, 
so therefore: 
\begin{equation}
  \label{equ:tointegrate}
 \int_{t=0}^{\delta t} \uu dt = - \MM^{-1} \cdot  \int_{t=0}^{\delta t} \frac{\partial\BB}{\partial t} dt.
\end{equation}
which can be differentiated with respect to time to yield equation~\ref{equ:velocity}.

\section{Nulls in a sum of two fields}

Magnetic fields bear characteristics that distinguish them from other fields in which topological defects are commonly studied.
Namely, magnetic fields take values in a vector space ($\mathbb{R}^3$) as opposed to some compact manifold ($S^2$ for ferromagnets, $\mathbb{R}P^2$ for nematic liquid crystals, etc.), with topological stability instead being enforced by a divergence-free condition.
Relatedly, sums of magnetic fields are well-defined, and magnetic fields are generally determined by non-local sources.
As a result, the \emph{local} dynamics of nulls is governed by \emph{global} conditions.
Here we show that this renders isotropes particularly useful in the modeling of dynamic magnetic field topologies.

Nulls are often studied where a local field is embedded in an external field, for example Earth's field embedded in the field generated by the Parker spiral~\citep{cowley1973qualitative}, the reversed field generated in an FRC~\citep{tuszewski1988field}, or nulls around a ring current as proposed for the Big Red Ball experiment~\citep{yu2022experimental}. 
Isotrope lines are a powerful tool to understand the location and movement of nulls in such configurations. 

Let us assume for simplicity that the external field is constant and in the direction $\hat{b}$.
For the local field, we know the magnitude and the isotropes. 
For the external and local fields to cancel at a point, the fields must be opposite in direction (i.e. the point lies on the isotrope corresponding with $-\hat{b}$) and equal in magnitude. 
This reduces the problem of locating nulls in three dimensions into locating the intersection of a curve (the isotrope) with a surface (of constant magnitude).

By a proper change of coordinates or choice of area form on $S^2$, this method of locating nulls may be extended to inhomogeneous externally sourced fields, so long as its field lines start and end on the domain boundary. Any such field can be used to construct a coordinate transform that makes the field constant within the bounds of the system. This is described in Appendix \ref{app:straightening}. In the resulting system, all techniques derived for homogeneous external fields may be applied before transforming back to the original system. Such a transformation is a diffeomorphism of the space and therefore preserves the topology of the magnetic field and its nulls. The isotrope field calculated after transformation will generally not be the transformation of the original isotrope field. Instead, it may be considered as the image of an isotrope field that implements a different area form on $S^2$.

\subsection{Motion and annihilation of nulls}
As the external field is varied smoothly, the location of a null changes smoothly.
If the strength is varied, nulls remain on their isotropes but move along them.
As the external field is strengthened (weakened) nulls will move up (down) the gradient of the local field's strength along the local field's isotrope.

The sum of the two fields has its own isotrope structure that can also be calculated to find the paths along which nulls will move as the externally applied field is varied.
The Jacobian $\MM$ can be calculated anywhere, and is independent of the external (and constant) field.
A null created at any location will will thus have the Jacobian (and it's topological index) determined by the properties of the local field.
As a result, the system is split into regions of $|\MM|\lessgtr 0$, where nulls of only one type or the other can appear.
These regions are separated by surfaces (or regions) on which $|\MM|=0$. 
We call these singular surfaces.
It can be shown that $\vv\cdot\nabla|\BB|=0$ on these surfaces.

Consider an isolated pair of nulls of opposite index.
As per figure~\ref{fig:attraction}, all but one isotrope leads from the index 1 null to the index -1 null. 
If an external field is added to this, both nulls move \emph{along the same isotrope}, either towards or away from each other (one parallel $\vv$ and one anti-parallel $\vv$). 
This leads us to formulate the concept of \emph{topological attraction}: Even if there is not an effective force acting on the nulls that moves them together, most any field that will move the nulls along an isotrope connecting them, until they merge.

As the strength of the external field is varied, if a null drifts onto a singular surface, it necessarily annihilates against a null of opposite type from the other side (following the opposite sign of $\vv\cdot\nabla|\BB|$).
This coincides with the breakdown of equation \ref{equ:velocity} as $\MM$ becomes non-invertible.
In this way, nulls may only be annihilated or created in pairs on singular surfaces of the Jacobian.

More generally, we can consider a pair of nulls merging due to a variation of both the direction and magnitude of a homogeneous external field.
Using the identities $\MM\cdot \vv=\BB \vv\cdot \nabla \ln|\BB|$ (globally) and $\vv\cdot \nabla |\BB|=0$ (at the singular surface), we see that $\MM\cdot \vv=0$ at the singular surface i. e. the isotrope field is a zero eigenvector of $\MM$ at the singular surface.
Eigenvectors and eigenvalues of $\MM$ vary smoothly in $R^3$, and $|\MM|$ changes sign across the singular surface.
Therefore, there is an eigenvalue that vanishes on the surface, whose eigenvector is $\sim v$ near the surface.
This eigenvector must be in the fan plane of both nulls since its eigenvalue changes sign between the nulls.
As two nulls merge at a singular surface, their fan planes therefore intersect along an isotrope connecting the two nulls.

\subsection{Nulls around a dipole}
The example of a planetary field (approximated as a dipole) in a locally homogeneous embedding field will be helpful to illustrate the above concepts. 
The nulls created in this configuration affect magnetic connectivity, and influence phenomena such as auroras on Earth \citep{tanaka2022interpretation}, and solar wind surface irradiation on other planets~\citep{winslow2012observations}. 
Nulls around a dipole have long been studied as a model, starting with~\cite{cowley1973qualitative} and continuing to this day~\citep{elderMagneticNullsInteracting2021}.

A magnetic dipole at the origin aligned with the z-axis has a direction
\begin{equation}\label{eq:dipole}
  \hat{\bb}\propto 3 \cos(\phi)\hat{r}-\hat{z}
\end{equation}
and magnitude
\begin{equation}\label{eq:dipolemag}
    |\mathbf{B}| = \frac{m}{r^3}\sqrt{1+3\cos^2(\phi)}
\end{equation}
where $\hat{r}=\xx/|\xx|$, and $\phi$ is the polar angle. 
The magnetic field direction only depends on $\theta$ and $\phi$, and is independent of $r$.
The components of the director map $g$ of the dipole (eq.~\eqref{eq:gfns}) can be written as:
\begin{equation}\label{equ:dipoleg}
\begin{split}
    g_{\theta}(r,\theta,\phi)=&
    \begin{cases}
        \theta & \phi<\pi/2\\
        \theta +\pi &\phi>\pi/2
    \end{cases}\\
    \arctan(g_{\phi}(r,\theta,\phi))=&\frac{3\sin(2\phi)}{1+3\cos(2\phi)}.
\end{split}
\end{equation}
As a result, the isotrope field is purely radial.
\begin{equation}
    \vv=-\frac{9 (7 \cos (\phi)+\cos (3 \phi))}{\sqrt{2} r^2 (3 \cos (2 \phi)+5)^{3/2}}\hat r.
\end{equation}


The nulls of a dipole in an external field can now be located by identifying the isotrope of opposite direction to the external field through equation~\ref{equ:dipoleg}, and finding the point where their magnitudes are equal through equation~\ref{eq:dipolemag}. 
For any external field there will be two nulls on opposite sides of the dipole. 
The nulls will have opposite index, which can be verified in two ways. 
The first is by using equation~\ref{eq:intdeg} on a surface $U$ of sufficiently large radius that the embedding field dominates. 
The degree of that map is zero since it maps all points on $U$ to the point on $S^2$ corresponding to the direction of the embedding field. 
The sum of the indices of the two enclosed nulls must therefore be zero. 
The second method is to evaluate $\MM$ of the dipole field, whose determinant is shown in Appendix \ref{app:dipole} to be negative for $z>0$, and positive for $z<0$. 

\begin{figure}
    \begin{overpic}[width=\linewidth]{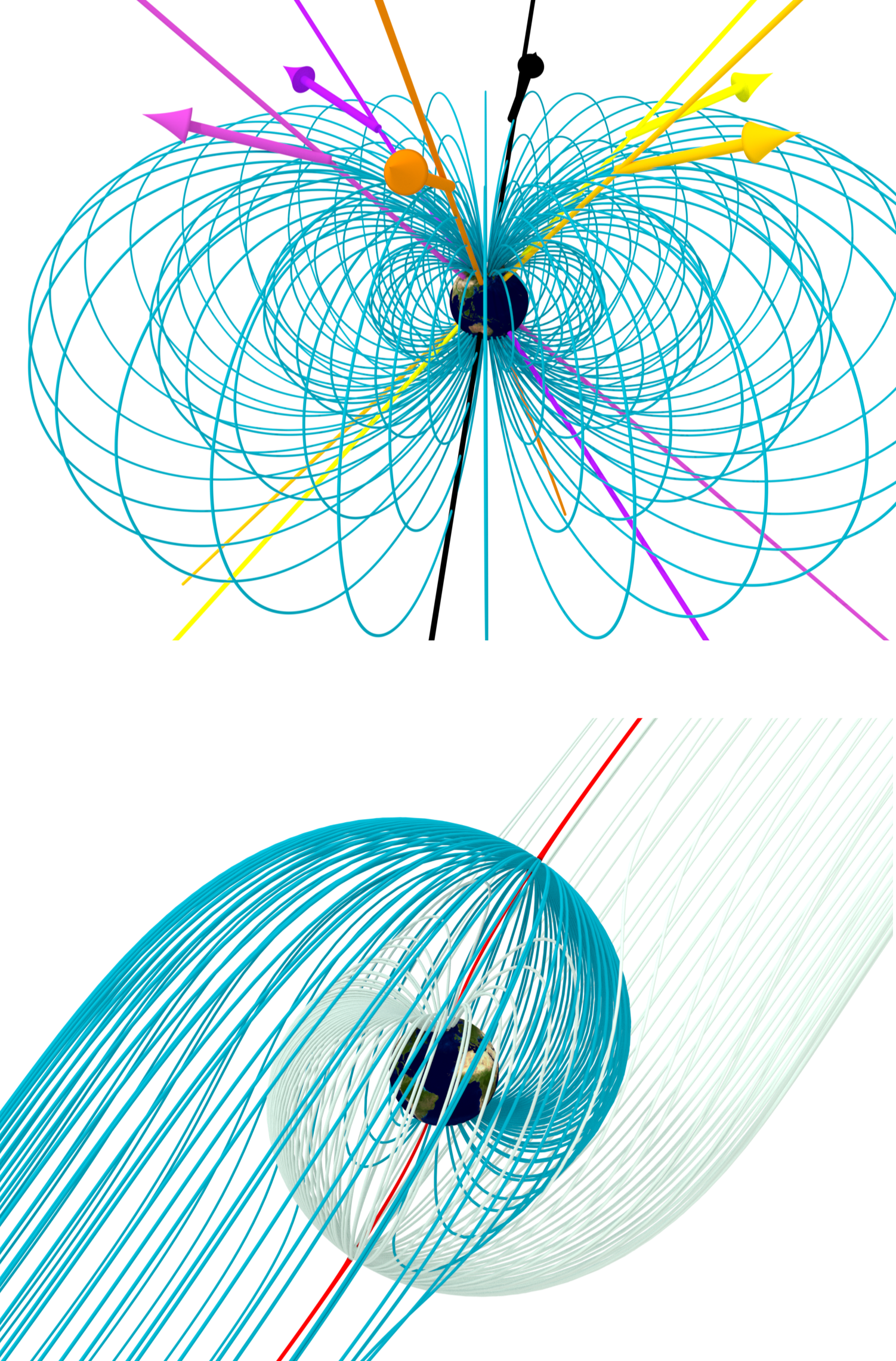}
    \put(1,95){\fcolorbox{white}{white}{\parbox{7pt}{\centering \textbf{a}}}}
    \put(1,45){\fcolorbox{white}{white}{\parbox{7pt}{\centering \textbf{b}}}}
    \end{overpic}
  \caption{Isotropes and nulls around the dipole.
  (a) The magnetic field and isotropes around a dipole field centered on a planet.
  The magnetic field lines are given by the blue curves, and seven isotropes are shown in different colors, with vectors indicating the magnetic field direction along the isotropes.
  (b) The magnetic field of a dipole embedded in an external field of direction $(-1,0,-1)$ (light yellow isotrope).
  The spines are colored red, the fan corresponding to the index $-1$ null is dark blue and the fan corresponding to the index $+1$ null is white.
  }
\end{figure}

For the purpose of laboratory astrophysics experiments, a dipole field may be approximated by that of a current loop. Such a geometry has been suggested for the study of magnetic reconnection in the vicinity of nulls in the Big Red Ball device at the Wisconsin Plasma Physics Laboratory.\citep{yu2022experimental}
For external fields weaker than the field strength at the center of the loop, the topology of the  field remains analogous to the above description.
For any field strength, isotropes may still be calculated in a closed albeit more complicated form involving elliptic integrals, and nulls may be located via an analogous calculation.

\subsection{Nulls around the Hopf fibration}
Another place where nulls may occur is a plasma containing a localized set of currents embedded in a surrounding magnetic field.
This is the case in magnetic clouds~\citep{smiet2019resistive}, FRC's~\citep{tuszewski1988field}, and compact toroids~\citep{degnan1993compact}. 
When the locally generated field is produced by a spatially extended current density (and not a singular point as in the dipole above), a much richer isotrope structure is produced.
A useful analytical model of such a local field is a vector field that lies tangent to fibers of the Hopf map~\citep{Hopf1931}. 
This configuration has been related to magnetic clouds~\citep{smiet2019resistive}, self-organizing structures in MHD~\citep{smiet2015self}, as the origin of galaxies~\citep{finkelstein1978magnetohydrodynamic}, and as topological MHD solitons~\citep{Kamchatnov1982}.
The geometry of the Hopf fibration also has many applications in other areas of physics~\citep{urbantke2003hopf}, including liquid crystals~\citep{chen2013generating}, solid-state physics~\citep{Dzyloshinskii1979}, optics~\cite{Irvine2008, Ranada1989}, and Bose-Einstein condensates~\citep{hall2016tying}.

The Hopf field is given by:
\begin{equation}
  \mathbf{B}_{\rm Hopf} = \frac{4}{\pi(1+r^2)^3}
            \begin{pmatrix}
              2(y- xz  ) \\
             -2(x + yz) \\
              -1+x^2+y^2-z^2
            \end{pmatrix}
\label{eq:hopffield}
\end{equation}
where $r=\sqrt{x^2+y^2+z^2}$.
See \citet{smiet2015self},~\citet{Kamchatnov1982}, or ~\citet{ranada1992em} for a derivation. 

The magnetic field strength of the Hopf field is given by the expression:
\begin{equation}\label{eq:hopfmag}
  |\mathbf{B}| = \frac{4}{\pi(1+r^2)^2}
\end{equation}
and surfaces of constant field strength are spheres centered on the origin.

\begin{figure}
  \includegraphics[width=0.5\textwidth]{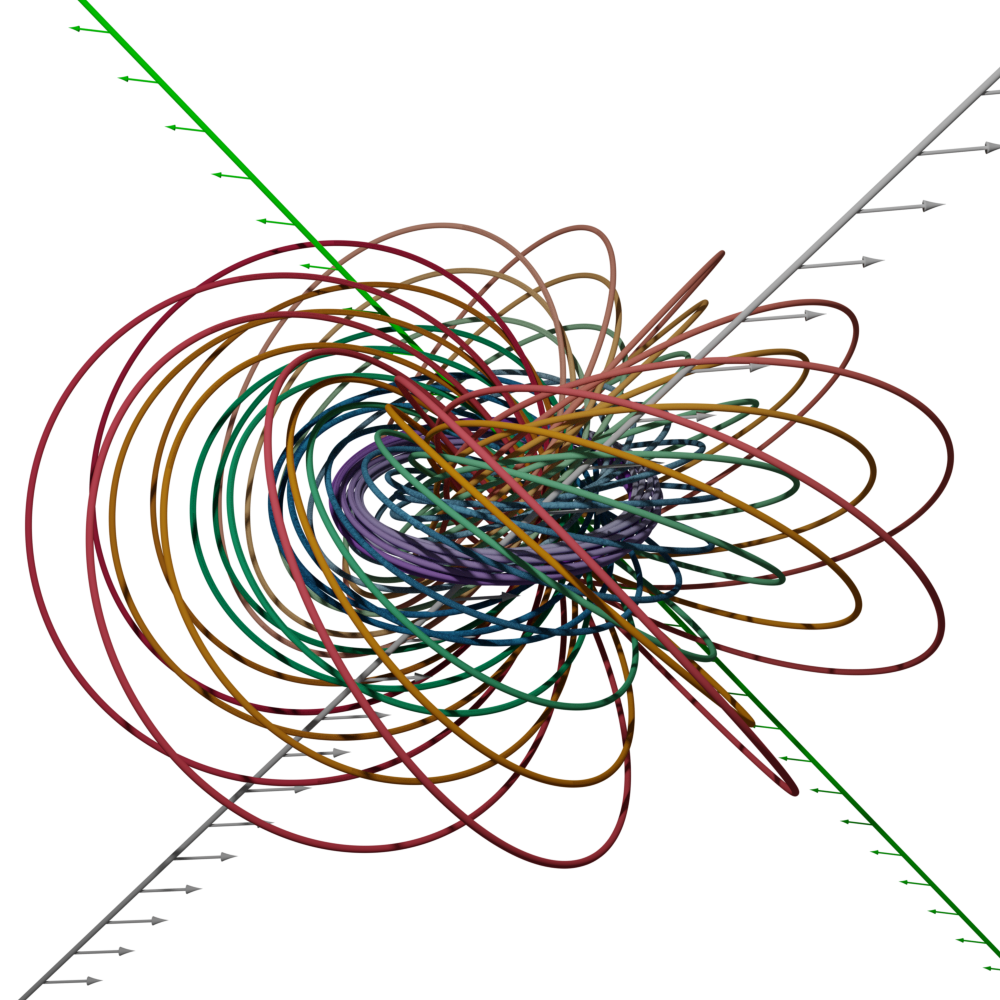}
  \caption{Field lines of the vector field given by equation~\eqref{eq:hopffield}. Every field line is a circle, and all field lines are linked with each other. Two isotropes, stream lines of equation~\eqref{eq:hopftrope}, are shown coresponding to the directions $\biso = ( 0.21 , -0.21 ,  0.03)$ (gray) and $-\biso$ (green).  
  }\label{fig:hopffield}
\end{figure}

We calculate the isotrope field of the Hopf field by inserting equation~\eqref{eq:hopffield} into equations~\eqref{eq:gfns} and~\eqref{eq:isotropes}. 
The isotrope field is given by:
\begin{equation}\label{eq:hopftrope}
  \vv = \frac{-4}{(1 +r^2)^2\left|-1 + x^2 + y^2 - z^2\right|}
  \begin{pmatrix}
    x z - y  \\
    x + y z \\
    1 + z^2 \\
\end{pmatrix}.
\end{equation}

Figure~\ref{fig:hopffield} shows select field lines of equation~\eqref{eq:hopffield}. 
The field consists purely of linked circles lying on a nested toroidal foliation. 
Two isotropes of the Hopf field, corresponding to $\biso  = ( 0.21 , -0.21 ,  0.03)$ (gray) and $-\biso$ (green) are also shown in figure~\ref{fig:hopffield}.
They  are integral curves of equation~\eqref{eq:hopftrope}, 
and the arrows along its length are in the direction of $\BB_{\rm Hopf}$.
The isotropes are straight lines which rule hyperoboid surfaces that form a foliation of $\mathbb{R}^3$.

We now can add an external field to the Hopf field, $\mathbf{B}_{\rm Hopf} +B_0 \biso$.
This will create two nulls on the isotrope corresponding with $-\biso$, i.e. the green isotrope in figure~\ref{fig:hopffield}. 
In the Hopf field we can study the merger of nulls, unlike in the dipole configuration, where all isotropes intersect at the singularity.

\begin{figure}
  \begin{overpic}[width=\linewidth]{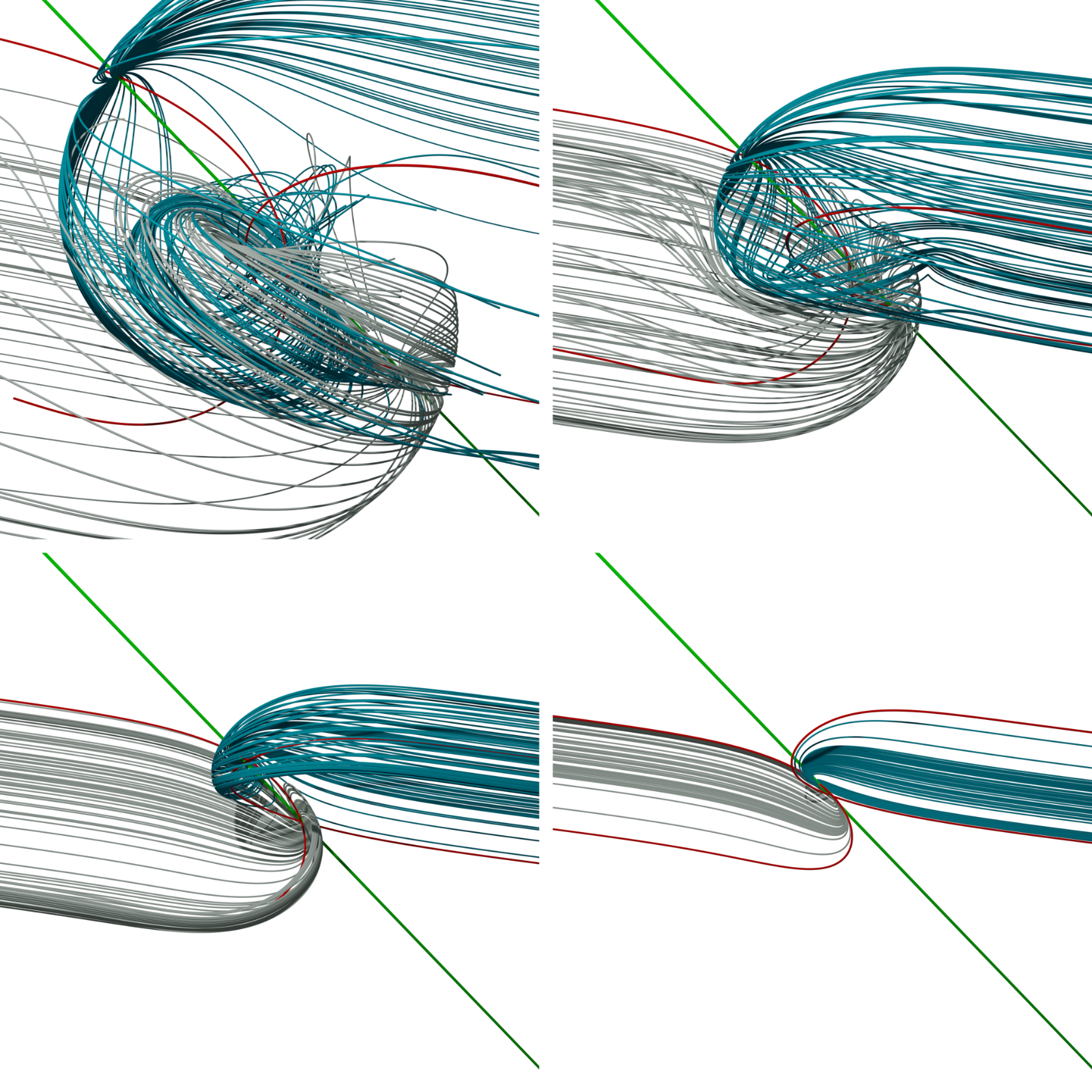}
    \put(1,95){\fcolorbox{white}{white}{\parbox{7pt}{\centering \textbf{a}}}}
    \put(51,95){\fcolorbox{white}{white}{\parbox{7pt}{\centering \textbf{b}}}}
    \put(1,45){\fcolorbox{white}{white}{\parbox{7pt}{\centering \textbf{c}}}}
    \put(51,45){\fcolorbox{white}{white}{\parbox{7pt}{\centering \textbf{d}}}}
  \end{overpic}
  \caption{Movement of nulls in the Hopf field plus a constant external field. The external field is given by $B_0 \biso$ with the value of $B_0$ (a) 0.01, (b) 0.065,  (c) 0.2, and (d) 0.3. The external field is opposite the field on the green isotrope in figure~\ref{fig:hopffield}, and thus the nulls move along it. As the magnitude is increased the two nulls move closer to each other, and merge on the $z=0$ plane. 
  We show the spines as red field lines and the fan planes by a series of blue (top null) and white (bottom null) field lines. 
  }\label{fig:hopfnulls}
\end{figure}
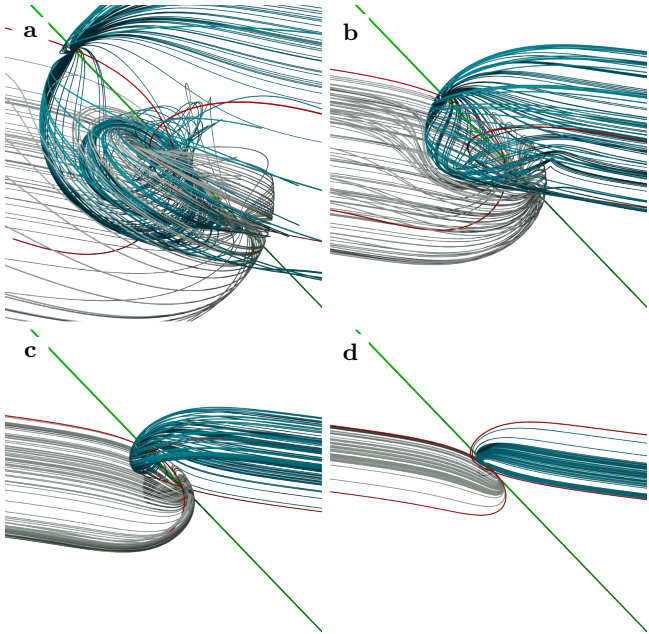

Figure~\ref{fig:hopfnulls} shows the movement of of the nulls as the magnetic field is increased from $B_0=0.01$ (a), to $0.065$ (b), $0.2$ (c), and 0.3 (d). 
As the external field is increased, a positive and a negative null appear from $z=+\infty$ and $z=-\infty$ respectively, and move along the isotrope (still pictured), until they merge on the $z=0$ plane when $B_0 = 0.31$.

Just as we calculated the isotropes of the Hopf field, we can also calculate the isotropes in this sum field. 
Instead of deriving an analytical expression for the isotropes, we evaluate the isotrope field numerically using automatic differentiation by implementing equation~\eqref{eq:isotropes} using the functional transformations in JAX~\citep{jax2018github}.

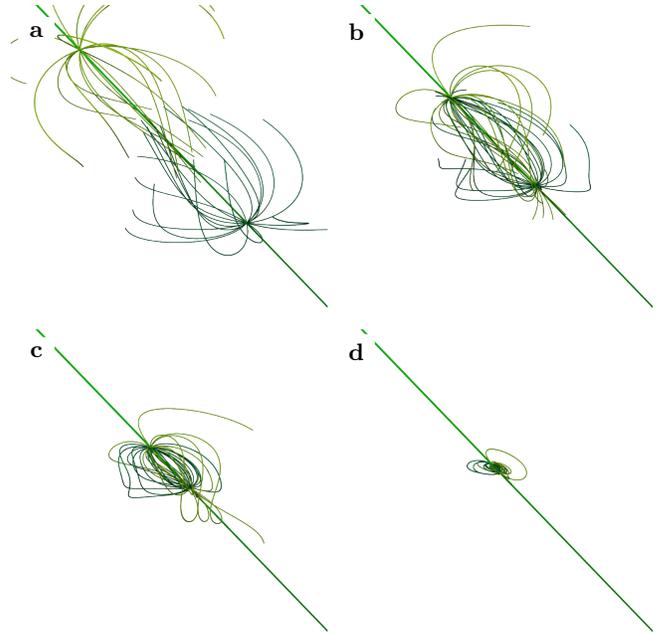
\begin{figure}
  \begin{overpic}[width=\linewidth]{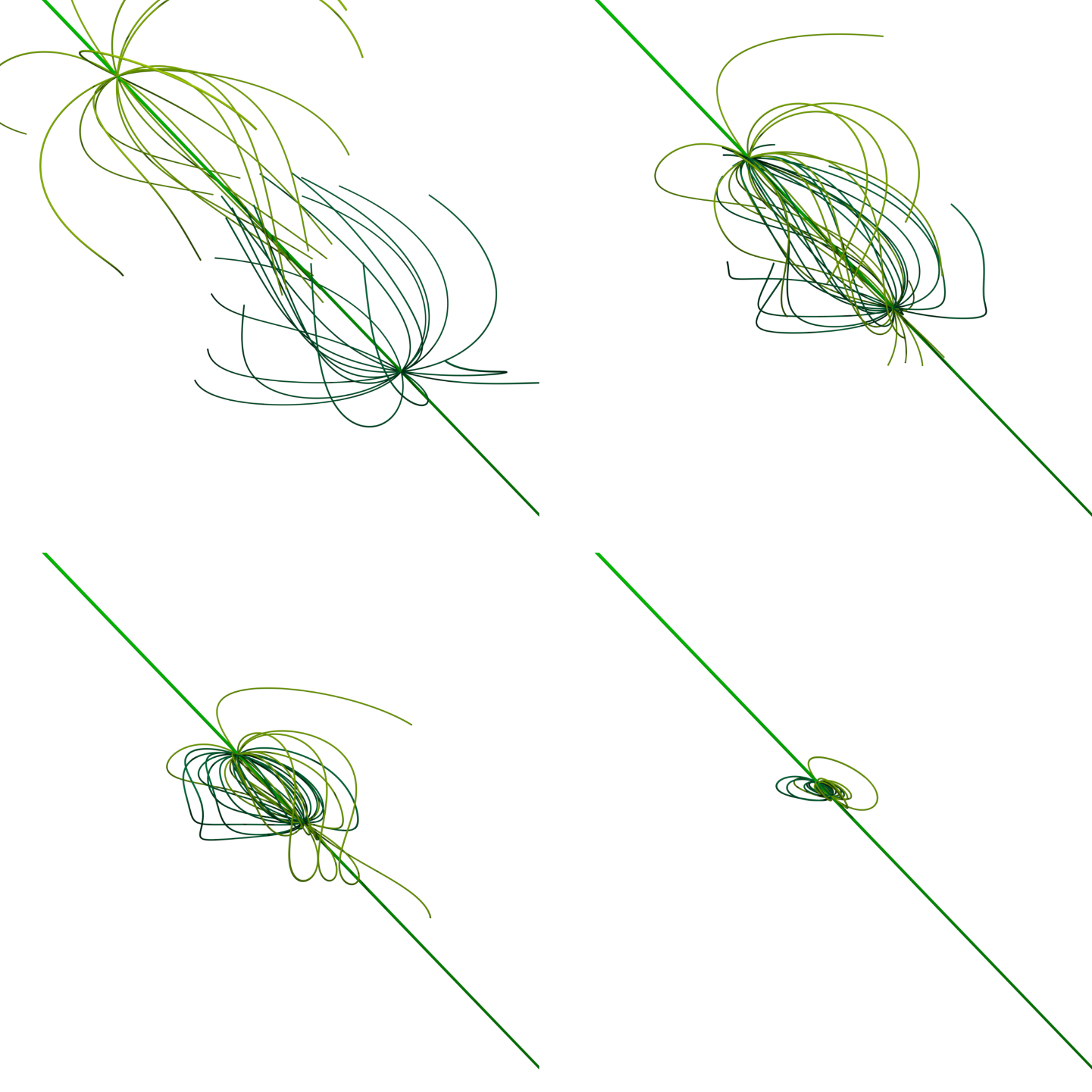}
    \put(1,95){\fcolorbox{white}{white}{\parbox{7pt}{\centering \textbf{a}}}}
    \put(51,95){\fcolorbox{white}{white}{\parbox{7pt}{\centering \textbf{b}}}}
    \put(1,45){\fcolorbox{white}{white}{\parbox{7pt}{\centering \textbf{c}}}}
    \put(51,45){\fcolorbox{white}{white}{\parbox{7pt}{\centering \textbf{d}}}}
  \end{overpic}
  \caption{Isotrope lines in the sum field $\mathbf{B}_{\rm Hopf} + B_0\biso$ during the merger of two nulls. Isotropes from the top null are light green and from the bottom null are dark green. The strength of the external field in panels (a)-(d) are identical to those in the corresponding panels in figure~\ref{fig:hopfnulls}.  The isotropes emerge from the top null, and converge on the bottom null. As the nulls converge, almost all the isotropes form a short path from one null to the other.
  }\label{fig:attraction}
\end{figure}

The isotrope field of the configurations in figure~\ref{fig:hopfnulls} are shown in figure~\ref{fig:attraction}. 
The top null is a source for the isotropes, and the bottom null is a sink. An isotrope corresponding with every direction will leave the positive null, and one of every direction will enter the negative null.
At infinity the Hopf field is zero, and the external field determines the direction of the field. 
Therefore, only the isotrope corresponding with the external field can go to infinity, all other isotropes connect the one null to the other, like the field lines around two opposite electric charges. 

\section{Conclusions and Discussion}
In this paper we have derived the concept of isotrope lines, lines in space where the magnetic field direction is constant. 
These lines are integral curves of the isotrope field, which is the Hodge dual of the pull-back of the area two-form on $S^2$ over the mapping defined by the direction of the magnetic field. 
As such, the isotropes convert the topological information of the mapping through which the topological index is defined into a local quantity that can be evaluated everywhere. 
Positive nulls are sources and negative nulls are sinks of this field, in a manner analogous to electrical point charges.
This construction bears some resemblance to analysis of topological defects in other fields, particularly in condensed matter.

We have used isotropes to analyze the location and movement of nulls in two toy model configurations relevant to astrophysics and plasma physics, namely the dipole and Hopf fields.
The nulls that appear are confined to specific isotropes corresponding to the direction opposite the external field.

We also analyzed the merger of two nulls in the Hopf field, both the magnetic structure in terms of the fan and spine planes of the nulls, and the isotrope structure during the merger. 
With isotropes we can also understand why opposite nulls are 'attracted' to each other, as they move along isotropes that connect them. 

Topological index also applies to higher order nulls, where the first order derivatives become zero, and an analysis in terms of $\MM$ breaks down. 
Progress has been made in describing such fields using second order derivatives~\citep{yang2017description, lukashenko2015general}, and the index provides a unifying framework to classify all nulls. 
Applying this work to higher order nulls remains for future work. 

There are many more applications of the concept of isotropes that did not fit in this paper. 
Isotropes in two dimensions can be used to locate and predict the movement of X- and O-points in Poincar\'e sections of magnetic fields in tokamaks and stellarators. 
In three dimensions, isotropes can be used to locate nulls in numerical simulations, where the isotrope field just needs to be integrated from any starting point to find a null, a process distinct from existing methods.\citep{olshevsky2020comparison,greene1992locating,haynesTrilinearMethodFinding2007,haynesMethodFindingThreedimensional2010}
Development and release of an algorithm for this is the subject of future work. 
Furthermore, the use of isotropes calculated via interpolation from spacecraft cluster data may offer a novel method for the location of magnetic nulls from this data.\citep{fuHowFindMagnetic2015,fu2020methods,guo3DReconnectionGeometries2022,heMagneticNullGeometry2008,olshevsky2020comparison}

\acknowledgements
We would like to thank Chaya Halperin for enlightening discussions which enabled us to connect this work to similar ideas in other fields. We would also like to thank Henry Fetsch and Yichen Fu for checking Appendices \ref{app:ndim} and \ref{app:straightening}, respectively, and Alexander Glasser for offering helpful suggestions concerning the phrasing of Appendix \ref{app:straightening}.

This work was supported by a grant from the Simons Foundation\ (1013657, JL).
This work was supported by the Excellence Center at WIS, the Simons grant 617006, the ISF grant 146845, and the NSF-BSF grant 2020765.
This work was supported by the U.S. Department of Energy under contract number DE-AC02-09CH11466. The United States Government retains a non-exclusive, paid-up, irrevocable, world-wide license to publish or reproduce the published form of this manuscript, or allow others to do so, for United States Government purposes.
This work has been carried out within the framework of the EUROfusion Consortium, partially funded by the European Union via the Euratom Research and Training Programme (Grant Agreement No 101052200 — EUROfusion). The Swiss contribution to this work has been funded by the Swiss State Secretariat for Education, Research and Innovation (SERI). Views and opinions expressed are however those of the author(s) only and do not necessarily reflect those of the European Union, the European Commission or SERI. Neither the European Union nor the European Commission nor SERI can be held responsible for them. 
BYI gratefully acknowledges financial support for this research by the Fulbright U.S. Scholar Program, which is sponsored by the U.S. Department of State and Fulbright Israel. Its contents are solely the responsibility of the author and do not necessarily represent the official views of the Fulbright Program, the Government of the United States, or Fulbright Israel.

\bibliography{refs}{}

\begin{thebibliography}{45}%
\makeatletter
\providecommand \@ifxundefined [1]{%
 \@ifx{#1\undefined}
}%
\providecommand \@ifnum [1]{%
 \ifnum #1\expandafter \@firstoftwo
 \else \expandafter \@secondoftwo
 \fi
}%
\providecommand \@ifx [1]{%
 \ifx #1\expandafter \@firstoftwo
 \else \expandafter \@secondoftwo
 \fi
}%
\providecommand \natexlab [1]{#1}%
\providecommand \enquote  [1]{``#1''}%
\providecommand \bibnamefont  [1]{#1}%
\providecommand \bibfnamefont [1]{#1}%
\providecommand \citenamefont [1]{#1}%
\providecommand \href@noop [0]{\@secondoftwo}%
\providecommand \href [0]{\begingroup \@sanitize@url \@href}%
\providecommand \@href[1]{\@@startlink{#1}\@@href}%
\providecommand \@@href[1]{\endgroup#1\@@endlink}%
\providecommand \@sanitize@url [0]{\catcode `\\12\catcode `\$12\catcode
  `\&12\catcode `\#12\catcode `\^12\catcode `\_12\catcode `\%12\relax}%
\providecommand \@@startlink[1]{}%
\providecommand \@@endlink[0]{}%
\providecommand \url  [0]{\begingroup\@sanitize@url \@url }%
\providecommand \@url [1]{\endgroup\@href {#1}{\urlprefix }}%
\providecommand \urlprefix  [0]{URL }%
\providecommand \Eprint [0]{\href }%
\providecommand \doibase [0]{https://doi.org/}%
\providecommand \selectlanguage [0]{\@gobble}%
\providecommand \bibinfo  [0]{\@secondoftwo}%
\providecommand \bibfield  [0]{\@secondoftwo}%
\providecommand \translation [1]{[#1]}%
\providecommand \BibitemOpen [0]{}%
\providecommand \bibitemStop [0]{}%
\providecommand \bibitemNoStop [0]{.\EOS\space}%
\providecommand \EOS [0]{\spacefactor3000\relax}%
\providecommand \BibitemShut  [1]{\csname bibitem#1\endcsname}%
\let\auto@bib@innerbib\@empty
\bibitem [{\citenamefont {Doyle}(1892)}]{doyle1979adventure}%
  \BibitemOpen
  \bibfield  {author} {\bibinfo {author} {\bibfnamefont {A.~C.}\ \bibnamefont
  {Doyle}},\ }\href@noop {} {\emph {\bibinfo {title} {The Adventure of Silver
  Blaze}}}\ (\bibinfo  {publisher} {Strand Magazine},\ \bibinfo {year}
  {1892})\BibitemShut {NoStop}%
\bibitem [{\citenamefont {Candelaresi}(2016)}]{candelaresi2016effects}%
  \BibitemOpen
  \bibfield  {author} {\bibinfo {author} {\bibfnamefont {S.}~\bibnamefont
  {Candelaresi}},\ }\bibfield  {title} {\bibinfo {title} {On the effects of
  magnetic field line topology on the energy propagation in the solar corona},\
  }\href@noop {} {\bibfield  {journal} {\bibinfo  {journal} {SPD}\ ,\ \bibinfo
  {pages} {401}} (\bibinfo {year} {2016})}\BibitemShut {NoStop}%
\bibitem [{\citenamefont {Priest}\ and\ \citenamefont
  {Titov}(1996)}]{priest1996magnetic}%
  \BibitemOpen
  \bibfield  {author} {\bibinfo {author} {\bibfnamefont {E.~R.}\ \bibnamefont
  {Priest}}\ and\ \bibinfo {author} {\bibfnamefont {V.}~\bibnamefont {Titov}},\
  }\bibfield  {title} {\bibinfo {title} {Magnetic reconnection at
  three-dimensional null points},\ }\href@noop {} {\bibfield  {journal}
  {\bibinfo  {journal} {Philosophical Transactions of the Royal Society of
  London. Series A: Mathematical, Physical and Engineering Sciences}\ }\textbf
  {\bibinfo {volume} {354}},\ \bibinfo {pages} {2951} (\bibinfo {year}
  {1996})}\BibitemShut {NoStop}%
\bibitem [{\citenamefont {Schindler}\ \emph {et~al.}(1988)\citenamefont
  {Schindler}, \citenamefont {Hesse},\ and\ \citenamefont
  {Birn}}]{schindler1988general}%
  \BibitemOpen
  \bibfield  {author} {\bibinfo {author} {\bibfnamefont {K.}~\bibnamefont
  {Schindler}}, \bibinfo {author} {\bibfnamefont {M.}~\bibnamefont {Hesse}},\
  and\ \bibinfo {author} {\bibfnamefont {J.}~\bibnamefont {Birn}},\ }\bibfield
  {title} {\bibinfo {title} {General magnetic reconnection, parallel electric
  fields, and helicity},\ }\href@noop {} {\bibfield  {journal} {\bibinfo
  {journal} {Journal of Geophysical Research: Space Physics}\ }\textbf
  {\bibinfo {volume} {93}},\ \bibinfo {pages} {5547} (\bibinfo {year}
  {1988})}\BibitemShut {NoStop}%
\bibitem [{\citenamefont {Greene}(1988)}]{greene1988geometrical}%
  \BibitemOpen
  \bibfield  {author} {\bibinfo {author} {\bibfnamefont {J.~M.}\ \bibnamefont
  {Greene}},\ }\bibfield  {title} {\bibinfo {title} {Geometrical properties of
  three-dimensional reconnecting magnetic fields with nulls},\ }\href@noop {}
  {\bibfield  {journal} {\bibinfo  {journal} {Journal of Geophysical Research:
  Space Physics (1978--2012)}\ }\textbf {\bibinfo {volume} {93}},\ \bibinfo
  {pages} {8583} (\bibinfo {year} {1988})}\BibitemShut {NoStop}%
\bibitem [{\citenamefont {Lau}\ and\ \citenamefont
  {Finn}(1990)}]{lau1990three}%
  \BibitemOpen
  \bibfield  {author} {\bibinfo {author} {\bibfnamefont {Y.-T.}\ \bibnamefont
  {Lau}}\ and\ \bibinfo {author} {\bibfnamefont {J.~M.}\ \bibnamefont {Finn}},\
  }\bibfield  {title} {\bibinfo {title} {Three-dimensional kinematic
  reconnection in the presence of field nulls and closed field lines},\
  }\href@noop {} {\bibfield  {journal} {\bibinfo  {journal} {The Astrophysical
  Journal}\ }\textbf {\bibinfo {volume} {350}},\ \bibinfo {pages} {672}
  (\bibinfo {year} {1990})}\BibitemShut {NoStop}%
\bibitem [{\citenamefont {Cowley}(1973)}]{cowley1973qualitative}%
  \BibitemOpen
  \bibfield  {author} {\bibinfo {author} {\bibfnamefont {S.}~\bibnamefont
  {Cowley}},\ }\bibfield  {title} {\bibinfo {title} {A qualitative study of the
  reconnection between the earth's magnetic field and an interplanetary field
  of arbitrary orientation},\ }\href@noop {} {\bibfield  {journal} {\bibinfo
  {journal} {Radio Science}\ }\textbf {\bibinfo {volume} {8}},\ \bibinfo
  {pages} {903} (\bibinfo {year} {1973})}\BibitemShut {NoStop}%
\bibitem [{\citenamefont {Tuszewski}(1988)}]{tuszewski1988field}%
  \BibitemOpen
  \bibfield  {author} {\bibinfo {author} {\bibfnamefont {M.}~\bibnamefont
  {Tuszewski}},\ }\bibfield  {title} {\bibinfo {title} {Field reversed
  configurations},\ }\href@noop {} {\bibfield  {journal} {\bibinfo  {journal}
  {Nuclear Fusion}\ }\textbf {\bibinfo {volume} {28}},\ \bibinfo {pages} {008}
  (\bibinfo {year} {1988})}\BibitemShut {NoStop}%
\bibitem [{\citenamefont {Parnell}\ \emph {et~al.}(1996)\citenamefont
  {Parnell}, \citenamefont {Smith}, \citenamefont {Neukirch},\ and\
  \citenamefont {Priest}}]{parnell1996structure}%
  \BibitemOpen
  \bibfield  {author} {\bibinfo {author} {\bibfnamefont {C.}~\bibnamefont
  {Parnell}}, \bibinfo {author} {\bibfnamefont {J.}~\bibnamefont {Smith}},
  \bibinfo {author} {\bibfnamefont {T.}~\bibnamefont {Neukirch}},\ and\
  \bibinfo {author} {\bibfnamefont {E.}~\bibnamefont {Priest}},\ }\bibfield
  {title} {\bibinfo {title} {The structure of three-dimensional magnetic
  neutral points},\ }\href@noop {} {\bibfield  {journal} {\bibinfo  {journal}
  {Physics of Plasmas (1994-present)}\ }\textbf {\bibinfo {volume} {3}},\
  \bibinfo {pages} {759} (\bibinfo {year} {1996})}\BibitemShut {NoStop}%
\bibitem [{\citenamefont {Greene}(1992)}]{greene1992locating}%
  \BibitemOpen
  \bibfield  {author} {\bibinfo {author} {\bibfnamefont {J.~M.}\ \bibnamefont
  {Greene}},\ }\bibfield  {title} {\bibinfo {title} {Locating three-dimensional
  roots by a bisection method},\ }\href@noop {} {\bibfield  {journal} {\bibinfo
   {journal} {Journal of Computational Physics}\ }\textbf {\bibinfo {volume}
  {98}},\ \bibinfo {pages} {194} (\bibinfo {year} {1992})}\BibitemShut
  {NoStop}%
\bibitem [{\citenamefont {Olshevsky}\ \emph {et~al.}(2020)\citenamefont
  {Olshevsky}, \citenamefont {Pontin}, \citenamefont {Williams}, \citenamefont
  {Parnell}, \citenamefont {Fu}, \citenamefont {Liu}, \citenamefont {Yao},\
  and\ \citenamefont {Khotyaintsev}}]{olshevsky2020comparison}%
  \BibitemOpen
  \bibfield  {author} {\bibinfo {author} {\bibfnamefont {V.}~\bibnamefont
  {Olshevsky}}, \bibinfo {author} {\bibfnamefont {D.}~\bibnamefont {Pontin}},
  \bibinfo {author} {\bibfnamefont {B.}~\bibnamefont {Williams}}, \bibinfo
  {author} {\bibfnamefont {C.}~\bibnamefont {Parnell}}, \bibinfo {author}
  {\bibfnamefont {H.}~\bibnamefont {Fu}}, \bibinfo {author} {\bibfnamefont
  {Y.}~\bibnamefont {Liu}}, \bibinfo {author} {\bibfnamefont {S.}~\bibnamefont
  {Yao}},\ and\ \bibinfo {author} {\bibfnamefont {Y.}~\bibnamefont
  {Khotyaintsev}},\ }\bibfield  {title} {\bibinfo {title} {A comparison of
  methods for finding magnetic nulls in simulations and in situ observations of
  space plasmas},\ }\href@noop {} {\bibfield  {journal} {\bibinfo  {journal}
  {Astronomy \& Astrophysics}\ }\textbf {\bibinfo {volume} {644}},\ \bibinfo
  {pages} {A150} (\bibinfo {year} {2020})}\BibitemShut {NoStop}%
\bibitem [{\citenamefont {Fu}\ \emph {et~al.}(2020)\citenamefont {Fu},
  \citenamefont {Wang}, \citenamefont {Zong}, \citenamefont {Chen},
  \citenamefont {He}, \citenamefont {Vaivads},\ and\ \citenamefont
  {Olshevsky}}]{fu2020methods}%
  \BibitemOpen
  \bibfield  {author} {\bibinfo {author} {\bibfnamefont {H.}~\bibnamefont
  {Fu}}, \bibinfo {author} {\bibfnamefont {Z.}~\bibnamefont {Wang}}, \bibinfo
  {author} {\bibfnamefont {Q.}~\bibnamefont {Zong}}, \bibinfo {author}
  {\bibfnamefont {X.}~\bibnamefont {Chen}}, \bibinfo {author} {\bibfnamefont
  {J.}~\bibnamefont {He}}, \bibinfo {author} {\bibfnamefont {A.}~\bibnamefont
  {Vaivads}},\ and\ \bibinfo {author} {\bibfnamefont {V.}~\bibnamefont
  {Olshevsky}},\ }\bibfield  {title} {\bibinfo {title} {Methods for finding
  magnetic nulls and reconstructing field topology: A review},\ }\href@noop {}
  {\bibfield  {journal} {\bibinfo  {journal} {Dayside Magnetosphere
  Interactions}\ ,\ \bibinfo {pages} {153}} (\bibinfo {year}
  {2020})}\BibitemShut {NoStop}%
\bibitem [{\citenamefont {Murphy}\ \emph {et~al.}(2015)\citenamefont {Murphy},
  \citenamefont {Parnell},\ and\ \citenamefont
  {Haynes}}]{murphy2015appearance}%
  \BibitemOpen
  \bibfield  {author} {\bibinfo {author} {\bibfnamefont {N.~A.}\ \bibnamefont
  {Murphy}}, \bibinfo {author} {\bibfnamefont {C.~E.}\ \bibnamefont
  {Parnell}},\ and\ \bibinfo {author} {\bibfnamefont {A.~L.}\ \bibnamefont
  {Haynes}},\ }\bibfield  {title} {\bibinfo {title} {The appearance, motion,
  and disappearance of three-dimensional magnetic null points},\ }\href@noop {}
  {\bibfield  {journal} {\bibinfo  {journal} {Physics of Plasmas}\ }\textbf
  {\bibinfo {volume} {22}},\ \bibinfo {pages} {102117} (\bibinfo {year}
  {2015})}\BibitemShut {NoStop}%
\bibitem [{\citenamefont
  {Flanders}(1963)}]{flandersVIApplicationsEuclidean1963}%
  \BibitemOpen
  \bibfield  {author} {\bibinfo {author} {\bibfnamefont {H.}~\bibnamefont
  {Flanders}},\ }\bibfield  {title} {\bibinfo {title} {{{VI Applications}} in
  {{Euclidean Space}}},\ }in\ \href@noop {} {\emph {\bibinfo {booktitle}
  {Differential Forms with Applications to the Physical Sciences}}},\ \bibinfo
  {series and number} {\bibinfo {series} {Mathematics in Science and
  Engineering}\ No.\ \bibinfo {number} {v. 11}}\ (\bibinfo  {publisher}
  {Academic Press},\ \bibinfo {address} {New York},\ \bibinfo {year}
  {1963})\BibitemShut {NoStop}%
\bibitem [{\citenamefont {Kurik}\ and\ \citenamefont
  {Lavrentovich}(1988)}]{kurikDefectsLiquidCrystals1988}%
  \BibitemOpen
  \bibfield  {author} {\bibinfo {author} {\bibfnamefont {M.~V.}\ \bibnamefont
  {Kurik}}\ and\ \bibinfo {author} {\bibfnamefont {O.~D.}\ \bibnamefont
  {Lavrentovich}},\ }\bibfield  {title} {\bibinfo {title} {Defects in liquid
  crystals: Homotopy theory and experimental studies},\ }\href
  {https://doi.org/10.1070/PU1988v031n03ABEH005710} {\bibfield  {journal}
  {\bibinfo  {journal} {Soviet Physics Uspekhi}\ }\textbf {\bibinfo {volume}
  {31}},\ \bibinfo {pages} {196} (\bibinfo {year} {1988})}\BibitemShut
  {NoStop}%
\bibitem [{\citenamefont {Alexander}\ \emph {et~al.}(2012)\citenamefont
  {Alexander}, \citenamefont {Chen}, \citenamefont {Matsumoto},\ and\
  \citenamefont {Kamien}}]{alexander2012nematic}%
  \BibitemOpen
  \bibfield  {author} {\bibinfo {author} {\bibfnamefont {G.~P.}\ \bibnamefont
  {Alexander}}, \bibinfo {author} {\bibfnamefont {B.~G.-g.}\ \bibnamefont
  {Chen}}, \bibinfo {author} {\bibfnamefont {E.~A.}\ \bibnamefont
  {Matsumoto}},\ and\ \bibinfo {author} {\bibfnamefont {R.~D.}\ \bibnamefont
  {Kamien}},\ }\bibfield  {title} {\bibinfo {title} {Colloquium: Disclination
  loops, point defects, and all that in nematic liquid crystals},\ }\href
  {https://doi.org/10.1103/RevModPhys.84.497} {\bibfield  {journal} {\bibinfo
  {journal} {Rev. Mod. Phys.}\ }\textbf {\bibinfo {volume} {84}},\ \bibinfo
  {pages} {497} (\bibinfo {year} {2012})}\BibitemShut {NoStop}%
\bibitem [{\citenamefont {Blaha}(1976)}]{blahaQuantizationRulesPoint1976}%
  \BibitemOpen
  \bibfield  {author} {\bibinfo {author} {\bibfnamefont {S.}~\bibnamefont
  {Blaha}},\ }\bibfield  {title} {\bibinfo {title} {Quantization {{Rules}} for
  {{Point Singularities}} in {{Superfluid He}} 3 and {{Liquid Crystals}}},\
  }\href {https://doi.org/10.1103/PhysRevLett.36.874} {\bibfield  {journal}
  {\bibinfo  {journal} {Physical Review Letters}\ }\textbf {\bibinfo {volume}
  {36}},\ \bibinfo {pages} {874} (\bibinfo {year} {1976})}\BibitemShut
  {NoStop}%
\bibitem [{\citenamefont {Trebin}(1982)}]{trebinTopologyNonuniformMedia1982}%
  \BibitemOpen
  \bibfield  {author} {\bibinfo {author} {\bibfnamefont {H.-R.}\ \bibnamefont
  {Trebin}},\ }\bibfield  {title} {\bibinfo {title} {The topology of
  non-uniform media in condensed matter physics},\ }\href
  {https://doi.org/10.1080/00018738200101458} {\bibfield  {journal} {\bibinfo
  {journal} {Advances in Physics}\ }\textbf {\bibinfo {volume} {31}},\ \bibinfo
  {pages} {195} (\bibinfo {year} {1982})}\BibitemShut {NoStop}%
\bibitem [{\citenamefont {De~Vega}(1978)}]{devegaClosedVorticesHopf1978}%
  \BibitemOpen
  \bibfield  {author} {\bibinfo {author} {\bibfnamefont {H.~J.}\ \bibnamefont
  {De~Vega}},\ }\bibfield  {title} {\bibinfo {title} {Closed vortices and the
  {{Hopf}} index in classical field theory},\ }\href
  {https://doi.org/10.1103/PhysRevD.18.2945} {\bibfield  {journal} {\bibinfo
  {journal} {Physical Review D}\ }\textbf {\bibinfo {volume} {18}},\ \bibinfo
  {pages} {2945} (\bibinfo {year} {1978})}\BibitemShut {NoStop}%
\bibitem [{\citenamefont {Smiet}\ \emph
  {et~al.}(2019{\natexlab{a}})\citenamefont {Smiet}, \citenamefont {Kramer},\
  and\ \citenamefont {Hudson}}]{smiet2019mapping}%
  \BibitemOpen
  \bibfield  {author} {\bibinfo {author} {\bibfnamefont {C.}~\bibnamefont
  {Smiet}}, \bibinfo {author} {\bibfnamefont {G.}~\bibnamefont {Kramer}},\ and\
  \bibinfo {author} {\bibfnamefont {S.}~\bibnamefont {Hudson}},\ }\bibfield
  {title} {\bibinfo {title} {Mapping the sawtooth},\ }\href@noop {} {\bibfield
  {journal} {\bibinfo  {journal} {Plasma Physics and Controlled Fusion}\
  }\textbf {\bibinfo {volume} {62}},\ \bibinfo {pages} {025007} (\bibinfo
  {year} {2019}{\natexlab{a}})}\BibitemShut {NoStop}%
\bibitem [{\citenamefont {Yu}\ and\ \citenamefont
  {Egedal}(2022)}]{yu2022experimental}%
  \BibitemOpen
  \bibfield  {author} {\bibinfo {author} {\bibfnamefont {X.}~\bibnamefont
  {Yu}}\ and\ \bibinfo {author} {\bibfnamefont {J.}~\bibnamefont {Egedal}},\
  }\bibfield  {title} {\bibinfo {title} {Experimental studies of magnetic
  reconnection in 3d geometries including magnetic nulls.},\ }\href@noop {}
  {\bibfield  {journal} {\bibinfo  {journal} {Bulletin of the American Physical
  Society}\ } (\bibinfo {year} {2022})}\BibitemShut {NoStop}%
\bibitem [{\citenamefont {Tanaka}\ \emph {et~al.}(2022)\citenamefont {Tanaka},
  \citenamefont {Ebihara}, \citenamefont {Watanabe}, \citenamefont {Fujita},
  \citenamefont {Nishitani},\ and\ \citenamefont
  {Kataoka}}]{tanaka2022interpretation}%
  \BibitemOpen
  \bibfield  {author} {\bibinfo {author} {\bibfnamefont {T.}~\bibnamefont
  {Tanaka}}, \bibinfo {author} {\bibfnamefont {Y.}~\bibnamefont {Ebihara}},
  \bibinfo {author} {\bibfnamefont {M.}~\bibnamefont {Watanabe}}, \bibinfo
  {author} {\bibfnamefont {S.}~\bibnamefont {Fujita}}, \bibinfo {author}
  {\bibfnamefont {N.}~\bibnamefont {Nishitani}},\ and\ \bibinfo {author}
  {\bibfnamefont {R.}~\bibnamefont {Kataoka}},\ }\bibfield  {title} {\bibinfo
  {title} {Interpretation of the theta aurora based on the null-separator
  structure},\ }\href@noop {} {\bibfield  {journal} {\bibinfo  {journal}
  {Journal of Geophysical Research: Space Physics}\ }\textbf {\bibinfo {volume}
  {127}},\ \bibinfo {pages} {e2022JA030332} (\bibinfo {year}
  {2022})}\BibitemShut {NoStop}%
\bibitem [{\citenamefont {Winslow}\ \emph {et~al.}(2012)\citenamefont
  {Winslow}, \citenamefont {Johnson}, \citenamefont {Anderson}, \citenamefont
  {Korth}, \citenamefont {Slavin}, \citenamefont {Purucker},\ and\
  \citenamefont {Solomon}}]{winslow2012observations}%
  \BibitemOpen
  \bibfield  {author} {\bibinfo {author} {\bibfnamefont {R.~M.}\ \bibnamefont
  {Winslow}}, \bibinfo {author} {\bibfnamefont {C.~L.}\ \bibnamefont
  {Johnson}}, \bibinfo {author} {\bibfnamefont {B.~J.}\ \bibnamefont
  {Anderson}}, \bibinfo {author} {\bibfnamefont {H.}~\bibnamefont {Korth}},
  \bibinfo {author} {\bibfnamefont {J.~A.}\ \bibnamefont {Slavin}}, \bibinfo
  {author} {\bibfnamefont {M.~E.}\ \bibnamefont {Purucker}},\ and\ \bibinfo
  {author} {\bibfnamefont {S.~C.}\ \bibnamefont {Solomon}},\ }\bibfield
  {title} {\bibinfo {title} {Observations of mercury's northern cusp region
  with messenger's magnetometer},\ }\href@noop {} {\bibfield  {journal}
  {\bibinfo  {journal} {Geophysical Research Letters}\ }\textbf {\bibinfo
  {volume} {39}} (\bibinfo {year} {2012})}\BibitemShut {NoStop}%
\bibitem [{\citenamefont {Elder}\ and\ \citenamefont
  {Boozer}(2021)}]{elderMagneticNullsInteracting2021}%
  \BibitemOpen
  \bibfield  {author} {\bibinfo {author} {\bibfnamefont {T.}~\bibnamefont
  {Elder}}\ and\ \bibinfo {author} {\bibfnamefont {A.~H.}\ \bibnamefont
  {Boozer}},\ }\bibfield  {title} {\bibinfo {title} {Magnetic nulls in
  interacting dipolar fields},\ }\href
  {https://doi.org/10.1017/S0022377821000210} {\bibfield  {journal} {\bibinfo
  {journal} {Journal of Plasma Physics}\ }\textbf {\bibinfo {volume} {87}},\
  \bibinfo {pages} {905870225} (\bibinfo {year} {2021})}\BibitemShut {NoStop}%
\bibitem [{\citenamefont {Smiet}\ \emph
  {et~al.}(2019{\natexlab{b}})\citenamefont {Smiet}, \citenamefont {de~Blank},
  \citenamefont {de~Jong}, \citenamefont {Kok},\ and\ \citenamefont
  {Bouwmeester}}]{smiet2019resistive}%
  \BibitemOpen
  \bibfield  {author} {\bibinfo {author} {\bibfnamefont {C.}~\bibnamefont
  {Smiet}}, \bibinfo {author} {\bibfnamefont {H.}~\bibnamefont {de~Blank}},
  \bibinfo {author} {\bibfnamefont {T.}~\bibnamefont {de~Jong}}, \bibinfo
  {author} {\bibfnamefont {D.}~\bibnamefont {Kok}},\ and\ \bibinfo {author}
  {\bibfnamefont {D.}~\bibnamefont {Bouwmeester}},\ }\bibfield  {title}
  {\bibinfo {title} {Resistive evolution of toroidal field distributions and
  their relation to magnetic clouds},\ }\href@noop {} {\bibfield  {journal}
  {\bibinfo  {journal} {Journal of Plasma Physics}\ }\textbf {\bibinfo {volume}
  {85}} (\bibinfo {year} {2019}{\natexlab{b}})}\BibitemShut {NoStop}%
\bibitem [{\citenamefont {Degnan}\ \emph {et~al.}(1993)\citenamefont {Degnan},
  \citenamefont {Peterkin~Jr}, \citenamefont {Baca}, \citenamefont {Beason},
  \citenamefont {Bell}, \citenamefont {Dearborn}, \citenamefont {Dietz},
  \citenamefont {Douglas}, \citenamefont {Englert}, \citenamefont {Englert}
  \emph {et~al.}}]{degnan1993compact}%
  \BibitemOpen
  \bibfield  {author} {\bibinfo {author} {\bibfnamefont {J.}~\bibnamefont
  {Degnan}}, \bibinfo {author} {\bibfnamefont {R.}~\bibnamefont {Peterkin~Jr}},
  \bibinfo {author} {\bibfnamefont {G.}~\bibnamefont {Baca}}, \bibinfo {author}
  {\bibfnamefont {J.}~\bibnamefont {Beason}}, \bibinfo {author} {\bibfnamefont
  {D.}~\bibnamefont {Bell}}, \bibinfo {author} {\bibfnamefont {M.}~\bibnamefont
  {Dearborn}}, \bibinfo {author} {\bibfnamefont {D.}~\bibnamefont {Dietz}},
  \bibinfo {author} {\bibfnamefont {M.}~\bibnamefont {Douglas}}, \bibinfo
  {author} {\bibfnamefont {S.}~\bibnamefont {Englert}}, \bibinfo {author}
  {\bibfnamefont {T.}~\bibnamefont {Englert}}, \emph {et~al.},\ }\bibfield
  {title} {\bibinfo {title} {Compact toroid formation, compression, and
  acceleration},\ }\href@noop {} {\bibfield  {journal} {\bibinfo  {journal}
  {Physics of Fluids B: Plasma Physics}\ }\textbf {\bibinfo {volume} {5}},\
  \bibinfo {pages} {2938} (\bibinfo {year} {1993})}\BibitemShut {NoStop}%
\bibitem [{\citenamefont {Hopf}(1931)}]{Hopf1931}%
  \BibitemOpen
  \bibfield  {author} {\bibinfo {author} {\bibfnamefont {H.}~\bibnamefont
  {Hopf}},\ }\bibfield  {title} {\bibinfo {title} {{\"{U}ber die Abbildungen
  der dreidimensionalen Sph\"{a}re auf die Kugelfl\"{a}che}},\ }\href@noop {}
  {\bibfield  {journal} {\bibinfo  {journal} {Math. Ann.}\ }\textbf {\bibinfo
  {volume} {104}},\ \bibinfo {pages} {637} (\bibinfo {year}
  {1931})}\BibitemShut {NoStop}%
\bibitem [{\citenamefont {Smiet}\ \emph {et~al.}(2015)\citenamefont {Smiet},
  \citenamefont {Candelaresi}, \citenamefont {Thompson}, \citenamefont
  {Swearngin}, \citenamefont {Dalhuisen},\ and\ \citenamefont
  {Bouwmeester}}]{smiet2015self}%
  \BibitemOpen
  \bibfield  {author} {\bibinfo {author} {\bibfnamefont {C.}~\bibnamefont
  {Smiet}}, \bibinfo {author} {\bibfnamefont {S.}~\bibnamefont {Candelaresi}},
  \bibinfo {author} {\bibfnamefont {A.}~\bibnamefont {Thompson}}, \bibinfo
  {author} {\bibfnamefont {J.}~\bibnamefont {Swearngin}}, \bibinfo {author}
  {\bibfnamefont {J.}~\bibnamefont {Dalhuisen}},\ and\ \bibinfo {author}
  {\bibfnamefont {D.}~\bibnamefont {Bouwmeester}},\ }\bibfield  {title}
  {\bibinfo {title} {Self-organizing knotted magnetic structures in plasma},\
  }\href@noop {} {\bibfield  {journal} {\bibinfo  {journal} {Physical review
  letters}\ }\textbf {\bibinfo {volume} {115}},\ \bibinfo {pages} {095001}
  (\bibinfo {year} {2015})}\BibitemShut {NoStop}%
\bibitem [{\citenamefont {Finkelstein}\ and\ \citenamefont
  {Weil}(1978)}]{finkelstein1978magnetohydrodynamic}%
  \BibitemOpen
  \bibfield  {author} {\bibinfo {author} {\bibfnamefont {D.}~\bibnamefont
  {Finkelstein}}\ and\ \bibinfo {author} {\bibfnamefont {D.}~\bibnamefont
  {Weil}},\ }\bibfield  {title} {\bibinfo {title} {Magnetohydrodynamic kinks in
  astrophysics},\ }\href@noop {} {\bibfield  {journal} {\bibinfo  {journal}
  {International Journal of Theoretical Physics}\ }\textbf {\bibinfo {volume}
  {17}},\ \bibinfo {pages} {201} (\bibinfo {year} {1978})}\BibitemShut
  {NoStop}%
\bibitem [{\citenamefont {Kamchatnov}(1982)}]{Kamchatnov1982}%
  \BibitemOpen
  \bibfield  {author} {\bibinfo {author} {\bibfnamefont {A.~M.}\ \bibnamefont
  {Kamchatnov}},\ }\bibfield  {title} {\bibinfo {title} {Topological solitons
  in magnetohydrodynamics},\ }\href@noop {} {\bibfield  {journal} {\bibinfo
  {journal} {Soviet Journal of Expermental and Theoretical Physics}\ }\textbf
  {\bibinfo {volume} {82}},\ \bibinfo {pages} {117} (\bibinfo {year}
  {1982})}\BibitemShut {NoStop}%
\bibitem [{\citenamefont {Urbantke}(2003)}]{urbantke2003hopf}%
  \BibitemOpen
  \bibfield  {author} {\bibinfo {author} {\bibfnamefont {H.}~\bibnamefont
  {Urbantke}},\ }\bibfield  {title} {\bibinfo {title} {The hopf fibration-seven
  times in physics},\ }\href@noop {} {\bibfield  {journal} {\bibinfo  {journal}
  {Journal of geometry and physics}\ }\textbf {\bibinfo {volume} {46}},\
  \bibinfo {pages} {125} (\bibinfo {year} {2003})}\BibitemShut {NoStop}%
\bibitem [{\citenamefont {Chen}\ \emph {et~al.}(2013)\citenamefont {Chen},
  \citenamefont {Ackerman}, \citenamefont {Alexander}, \citenamefont {Kamien},\
  and\ \citenamefont {Smalyukh}}]{chen2013generating}%
  \BibitemOpen
  \bibfield  {author} {\bibinfo {author} {\bibfnamefont {B.~G.-g.}\
  \bibnamefont {Chen}}, \bibinfo {author} {\bibfnamefont {P.~J.}\ \bibnamefont
  {Ackerman}}, \bibinfo {author} {\bibfnamefont {G.~P.}\ \bibnamefont
  {Alexander}}, \bibinfo {author} {\bibfnamefont {R.~D.}\ \bibnamefont
  {Kamien}},\ and\ \bibinfo {author} {\bibfnamefont {I.~I.}\ \bibnamefont
  {Smalyukh}},\ }\bibfield  {title} {\bibinfo {title} {Generating the hopf
  fibration experimentally in nematic liquid crystals},\ }\href@noop {}
  {\bibfield  {journal} {\bibinfo  {journal} {Physical review letters}\
  }\textbf {\bibinfo {volume} {110}},\ \bibinfo {pages} {237801} (\bibinfo
  {year} {2013})}\BibitemShut {NoStop}%
\bibitem [{\citenamefont {Dzyloshinskii}\ and\ \citenamefont
  {Ivanov}(1979)}]{Dzyloshinskii1979}%
  \BibitemOpen
  \bibfield  {author} {\bibinfo {author} {\bibfnamefont {I.~E.}\ \bibnamefont
  {Dzyloshinskii}}\ and\ \bibinfo {author} {\bibfnamefont {B.~A.}\ \bibnamefont
  {Ivanov}},\ }\bibfield  {title} {\bibinfo {title} {Localized topological
  solitons in a ferromagnet},\ }\href@noop {} {\bibfield  {journal} {\bibinfo
  {journal} {Journal of Experimental and Theoretical Physics}\ }\textbf
  {\bibinfo {volume} {29}},\ \bibinfo {pages} {540} (\bibinfo {year}
  {1979})}\BibitemShut {NoStop}%
\bibitem [{\citenamefont {Irvine}\ and\ \citenamefont
  {Bouwmeester}(2008)}]{Irvine2008}%
  \BibitemOpen
  \bibfield  {author} {\bibinfo {author} {\bibfnamefont {W.~T.~M.}\
  \bibnamefont {Irvine}}\ and\ \bibinfo {author} {\bibfnamefont
  {D.}~\bibnamefont {Bouwmeester}},\ }\bibfield  {title} {\bibinfo {title}
  {Linked and knotted beams of light},\ }\href@noop {} {\bibfield  {journal}
  {\bibinfo  {journal} {Nature Physics}\ }\textbf {\bibinfo {volume} {4}},\
  \bibinfo {pages} {716} (\bibinfo {year} {2008})}\BibitemShut {NoStop}%
\bibitem [{\citenamefont {Ra{\~n}ada}(1989)}]{Ranada1989}%
  \BibitemOpen
  \bibfield  {author} {\bibinfo {author} {\bibfnamefont {A.~F.}\ \bibnamefont
  {Ra{\~n}ada}},\ }\bibfield  {title} {\bibinfo {title} {A topological theory
  of the electromagnetic field},\ }\href@noop {} {\bibfield  {journal}
  {\bibinfo  {journal} {Letters in Mathematical Physics}\ }\textbf {\bibinfo
  {volume} {18}},\ \bibinfo {pages} {97} (\bibinfo {year} {1989})}\BibitemShut
  {NoStop}%
\bibitem [{\citenamefont {Hall}\ \emph {et~al.}(2016)\citenamefont {Hall},
  \citenamefont {Ray}, \citenamefont {Tiurev}, \citenamefont {Ruokokoski},
  \citenamefont {Gheorghe},\ and\ \citenamefont
  {M{\"o}tt{\"o}nen}}]{hall2016tying}%
  \BibitemOpen
  \bibfield  {author} {\bibinfo {author} {\bibfnamefont {D.~S.}\ \bibnamefont
  {Hall}}, \bibinfo {author} {\bibfnamefont {M.~W.}\ \bibnamefont {Ray}},
  \bibinfo {author} {\bibfnamefont {K.}~\bibnamefont {Tiurev}}, \bibinfo
  {author} {\bibfnamefont {E.}~\bibnamefont {Ruokokoski}}, \bibinfo {author}
  {\bibfnamefont {A.~H.}\ \bibnamefont {Gheorghe}},\ and\ \bibinfo {author}
  {\bibfnamefont {M.}~\bibnamefont {M{\"o}tt{\"o}nen}},\ }\bibfield  {title}
  {\bibinfo {title} {Tying quantum knots},\ }\href@noop {} {\bibfield
  {journal} {\bibinfo  {journal} {Nature Physics}\ } (\bibinfo {year}
  {2016})}\BibitemShut {NoStop}%
\bibitem [{\citenamefont {Ranada}(1992)}]{ranada1992em}%
  \BibitemOpen
  \bibfield  {author} {\bibinfo {author} {\bibfnamefont {A.~F.}\ \bibnamefont
  {Ranada}},\ }\bibfield  {title} {\bibinfo {title} {Topological
  electromagnetism},\ }\href {https://doi.org/10.1088/0305-4470/25/6/020}
  {\bibfield  {journal} {\bibinfo  {journal} {Journal of Physics A:
  Mathematical and General}\ }\textbf {\bibinfo {volume} {25}},\ \bibinfo
  {pages} {1621} (\bibinfo {year} {1992})}\BibitemShut {NoStop}%
\bibitem [{\citenamefont {Bradbury}\ \emph {et~al.}(2018)\citenamefont
  {Bradbury}, \citenamefont {Frostig}, \citenamefont {Hawkins}, \citenamefont
  {Johnson}, \citenamefont {Leary}, \citenamefont {Maclaurin}, \citenamefont
  {Necula}, \citenamefont {Paszke}, \citenamefont {Vander{P}las}, \citenamefont
  {Wanderman-{M}ilne},\ and\ \citenamefont {Zhang}}]{jax2018github}%
  \BibitemOpen
  \bibfield  {author} {\bibinfo {author} {\bibfnamefont {J.}~\bibnamefont
  {Bradbury}}, \bibinfo {author} {\bibfnamefont {R.}~\bibnamefont {Frostig}},
  \bibinfo {author} {\bibfnamefont {P.}~\bibnamefont {Hawkins}}, \bibinfo
  {author} {\bibfnamefont {M.~J.}\ \bibnamefont {Johnson}}, \bibinfo {author}
  {\bibfnamefont {C.}~\bibnamefont {Leary}}, \bibinfo {author} {\bibfnamefont
  {D.}~\bibnamefont {Maclaurin}}, \bibinfo {author} {\bibfnamefont
  {G.}~\bibnamefont {Necula}}, \bibinfo {author} {\bibfnamefont
  {A.}~\bibnamefont {Paszke}}, \bibinfo {author} {\bibfnamefont
  {J.}~\bibnamefont {Vander{P}las}}, \bibinfo {author} {\bibfnamefont
  {S.}~\bibnamefont {Wanderman-{M}ilne}},\ and\ \bibinfo {author}
  {\bibfnamefont {Q.}~\bibnamefont {Zhang}},\ }\href
  {http://github.com/google/jax} {\bibinfo {title} {{JAX}: composable
  transformations of {P}ython+{N}um{P}y programs}} (\bibinfo {year}
  {2018})\BibitemShut {NoStop}%
\bibitem [{\citenamefont {Yang}(2017)}]{yang2017description}%
  \BibitemOpen
  \bibfield  {author} {\bibinfo {author} {\bibfnamefont {S.-D.}\ \bibnamefont
  {Yang}},\ }\bibfield  {title} {\bibinfo {title} {Description of second-order
  three-dimensional magnetic neutral points},\ }\href@noop {} {\bibfield
  {journal} {\bibinfo  {journal} {Physics of Plasmas}\ }\textbf {\bibinfo
  {volume} {24}},\ \bibinfo {pages} {012903} (\bibinfo {year}
  {2017})}\BibitemShut {NoStop}%
\bibitem [{\citenamefont {Lukashenko}\ and\ \citenamefont
  {Veselovsky}(2015)}]{lukashenko2015general}%
  \BibitemOpen
  \bibfield  {author} {\bibinfo {author} {\bibfnamefont {A.}~\bibnamefont
  {Lukashenko}}\ and\ \bibinfo {author} {\bibfnamefont {I.}~\bibnamefont
  {Veselovsky}},\ }\bibfield  {title} {\bibinfo {title} {General principles of
  describing second-and higher-order null points of a potential magnetic field
  in 3d},\ }\href@noop {} {\bibfield  {journal} {\bibinfo  {journal}
  {Geomagnetism and Aeronomy}\ }\textbf {\bibinfo {volume} {55}},\ \bibinfo
  {pages} {1152} (\bibinfo {year} {2015})}\BibitemShut {NoStop}%
\bibitem [{\citenamefont {Haynes}\ and\ \citenamefont
  {Parnell}(2007)}]{haynesTrilinearMethodFinding2007}%
  \BibitemOpen
  \bibfield  {author} {\bibinfo {author} {\bibfnamefont {A.~L.}\ \bibnamefont
  {Haynes}}\ and\ \bibinfo {author} {\bibfnamefont {C.~E.}\ \bibnamefont
  {Parnell}},\ }\bibfield  {title} {\bibinfo {title} {A trilinear method for
  finding null points in a three-dimensional vector space},\ }\href
  {https://doi.org/10.1063/1.2756751} {\bibfield  {journal} {\bibinfo
  {journal} {Physics of Plasmas}\ }\textbf {\bibinfo {volume} {14}},\ \bibinfo
  {pages} {082107} (\bibinfo {year} {2007})}\BibitemShut {NoStop}%
\bibitem [{\citenamefont {Haynes}\ and\ \citenamefont
  {Parnell}(2010)}]{haynesMethodFindingThreedimensional2010}%
  \BibitemOpen
  \bibfield  {author} {\bibinfo {author} {\bibfnamefont {A.~L.}\ \bibnamefont
  {Haynes}}\ and\ \bibinfo {author} {\bibfnamefont {C.~E.}\ \bibnamefont
  {Parnell}},\ }\bibfield  {title} {\bibinfo {title} {A method for finding
  three-dimensional magnetic skeletons},\ }\href
  {https://doi.org/10.1063/1.3467499} {\bibfield  {journal} {\bibinfo
  {journal} {Physics of Plasmas}\ }\textbf {\bibinfo {volume} {17}},\ \bibinfo
  {pages} {092903} (\bibinfo {year} {2010})}\BibitemShut {NoStop}%
\bibitem [{\citenamefont {Fu}\ \emph {et~al.}(2015)\citenamefont {Fu},
  \citenamefont {Vaivads}, \citenamefont {Khotyaintsev}, \citenamefont
  {Olshevsky}, \citenamefont {Andr{\'e}}, \citenamefont {Cao}, \citenamefont
  {Huang}, \citenamefont {Retin{\`o}},\ and\ \citenamefont
  {Lapenta}}]{fuHowFindMagnetic2015}%
  \BibitemOpen
  \bibfield  {author} {\bibinfo {author} {\bibfnamefont {H.~S.}\ \bibnamefont
  {Fu}}, \bibinfo {author} {\bibfnamefont {A.}~\bibnamefont {Vaivads}},
  \bibinfo {author} {\bibfnamefont {Y.~V.}\ \bibnamefont {Khotyaintsev}},
  \bibinfo {author} {\bibfnamefont {V.}~\bibnamefont {Olshevsky}}, \bibinfo
  {author} {\bibfnamefont {M.}~\bibnamefont {Andr{\'e}}}, \bibinfo {author}
  {\bibfnamefont {J.~B.}\ \bibnamefont {Cao}}, \bibinfo {author} {\bibfnamefont
  {S.~Y.}\ \bibnamefont {Huang}}, \bibinfo {author} {\bibfnamefont
  {A.}~\bibnamefont {Retin{\`o}}},\ and\ \bibinfo {author} {\bibfnamefont
  {G.}~\bibnamefont {Lapenta}},\ }\bibfield  {title} {\bibinfo {title} {How to
  find magnetic nulls and reconstruct field topology with {{MMS}} data?},\
  }\href {https://doi.org/10.1002/2015JA021082} {\bibfield  {journal} {\bibinfo
   {journal} {Journal of Geophysical Research: Space Physics}\ }\textbf
  {\bibinfo {volume} {120}},\ \bibinfo {pages} {3758} (\bibinfo {year}
  {2015})}\BibitemShut {NoStop}%
\bibitem [{\citenamefont {Guo}\ \emph {et~al.}(2022)\citenamefont {Guo},
  \citenamefont {Pu}, \citenamefont {Wang}, \citenamefont {Xiao},\ and\
  \citenamefont {He}}]{guo3DReconnectionGeometries2022}%
  \BibitemOpen
  \bibfield  {author} {\bibinfo {author} {\bibfnamefont {R.}~\bibnamefont
  {Guo}}, \bibinfo {author} {\bibfnamefont {Z.}~\bibnamefont {Pu}}, \bibinfo
  {author} {\bibfnamefont {X.}~\bibnamefont {Wang}}, \bibinfo {author}
  {\bibfnamefont {C.}~\bibnamefont {Xiao}},\ and\ \bibinfo {author}
  {\bibfnamefont {J.}~\bibnamefont {He}},\ }\bibfield  {title} {\bibinfo
  {title} {{{3D Reconnection Geometries With Magnetic Nulls}}:
  {{Multispacecraft Observations}} and {{Reconstructions}}},\ }\href
  {https://doi.org/10.1029/2021JA030248} {\bibfield  {journal} {\bibinfo
  {journal} {Journal of Geophysical Research: Space Physics}\ }\textbf
  {\bibinfo {volume} {127}},\ \bibinfo {pages} {e2021JA030248} (\bibinfo {year}
  {2022})}\BibitemShut {NoStop}%
\bibitem [{\citenamefont {He}\ \emph {et~al.}(2008)\citenamefont {He},
  \citenamefont {Tu}, \citenamefont {Tian}, \citenamefont {Xiao}, \citenamefont
  {Wang}, \citenamefont {Pu}, \citenamefont {Ma}, \citenamefont {Dunlop},
  \citenamefont {Zhao}, \citenamefont {Zhou}, \citenamefont {Wang},
  \citenamefont {Fu}, \citenamefont {Liu}, \citenamefont {Zong}, \citenamefont
  {Glassmeier}, \citenamefont {Reme}, \citenamefont {Dandouras},\ and\
  \citenamefont {Escoubet}}]{heMagneticNullGeometry2008}%
  \BibitemOpen
  \bibfield  {author} {\bibinfo {author} {\bibfnamefont {J.-S.}\ \bibnamefont
  {He}}, \bibinfo {author} {\bibfnamefont {C.-Y.}\ \bibnamefont {Tu}}, \bibinfo
  {author} {\bibfnamefont {H.}~\bibnamefont {Tian}}, \bibinfo {author}
  {\bibfnamefont {C.-J.}\ \bibnamefont {Xiao}}, \bibinfo {author}
  {\bibfnamefont {X.-G.}\ \bibnamefont {Wang}}, \bibinfo {author}
  {\bibfnamefont {Z.-Y.}\ \bibnamefont {Pu}}, \bibinfo {author} {\bibfnamefont
  {Z.-W.}\ \bibnamefont {Ma}}, \bibinfo {author} {\bibfnamefont {M.~W.}\
  \bibnamefont {Dunlop}}, \bibinfo {author} {\bibfnamefont {H.}~\bibnamefont
  {Zhao}}, \bibinfo {author} {\bibfnamefont {G.-P.}\ \bibnamefont {Zhou}},
  \bibinfo {author} {\bibfnamefont {J.-X.}\ \bibnamefont {Wang}}, \bibinfo
  {author} {\bibfnamefont {S.-Y.}\ \bibnamefont {Fu}}, \bibinfo {author}
  {\bibfnamefont {Z.-X.}\ \bibnamefont {Liu}}, \bibinfo {author} {\bibfnamefont
  {Q.-G.}\ \bibnamefont {Zong}}, \bibinfo {author} {\bibfnamefont {K.-H.}\
  \bibnamefont {Glassmeier}}, \bibinfo {author} {\bibfnamefont
  {H.}~\bibnamefont {Reme}}, \bibinfo {author} {\bibfnamefont {I.}~\bibnamefont
  {Dandouras}},\ and\ \bibinfo {author} {\bibfnamefont {C.~P.}\ \bibnamefont
  {Escoubet}},\ }\bibfield  {title} {\bibinfo {title} {A magnetic null geometry
  reconstructed from {{Cluster}} spacecraft observations},\ }\bibfield
  {journal} {\bibinfo  {journal} {Journal of Geophysical Research: Space
  Physics}\ }\textbf {\bibinfo {volume} {113}},\ \href
  {https://doi.org/10.1029/2007JA012609} {10.1029/2007JA012609} (\bibinfo
  {year} {2008})\BibitemShut {NoStop}%
\end{thebibliography}%

\clearpage
\onecolumngrid
\appendix
\pagenumbering{alph}

\section{Equivalence of expressions for the isotrope field}
\label{app:lit}
Here we demonstrate that our expression for the isotrope field in three dimensions, equation \ref{eq:isotropefield}, is equivalent to equation \ref{eq:litisotrope} found in work concerning ferromagnets (with the assumption of a uniform metric on $S^2$)~\citep{trebinTopologyNonuniformMedia1982}.
Similar expressions may be found in work concerning superfluids, liquid crystals, and topological field theories \citep{blahaQuantizationRulesPoint1976,devegaClosedVorticesHopf1978}.
The expression may be written:
\begin{equation}
    \vv=\frac{1}{2} \varepsilon_{i j k} \hat{b}_i\left(\nabla \hat{b}_j \times \nabla \hat{b}_k\right).
\end{equation}
We write the unit magnetic field vector as
\begin{equation}
    \hat{\mathbf{b}}(\mathbf{x})=h(g(\mathbf{x}))
\end{equation}
where we specify that $g(\mathbf{x})$ is the function defined by equation \ref{eq:director} from real space $\mathbb{R}^3$ to coordinates $(\alpha,\ \beta)$ on the unit sphere, and $h(\alpha,\beta)$ is the function mapping from these coordinates on the unit sphere to the cartesian coordinates $(h_x,h_y,h_z)$ of the standard embedding of the unit sphere in $\mathbb{R}^3$. In terms of these functions, $v$ may then be written:
\begin{equation}
    v_a=\frac{1}{2} \varepsilon_{i j k} \varepsilon_{a b c} h_i\left(\frac{\partial h_j}{\partial \alpha} \frac{\partial g_\alpha}{\partial x_b}+\frac{\partial h_j}{\partial \beta} \frac{\partial g_\beta}{\partial x_b}\right)\left(\frac{\partial h_k}{\partial \alpha} \frac{\partial g_\alpha}{\partial x_c}+\frac{\partial h_k}{\partial \beta} \frac{\partial g_\beta}{\partial x_c}\right).
\end{equation}
Making use of the identity $\varepsilon_{a b c} \frac{\partial g_\alpha}{\partial x_b} \frac{\partial g_\alpha}{\partial x_c}=\varepsilon_{a b c} \frac{\partial g_\beta}{\partial x_b} \frac{\partial g_\beta}{\partial x_c}=0$ and permuting indices yields
\begin{equation}
    v_a=\varepsilon_{i j k} \varepsilon_{a b c} h_i \frac{\partial h_j}{\partial \alpha} \frac{\partial g_\alpha}{\partial x_b} \frac{\partial h_k}{\partial \beta} \frac{\partial g_\beta}{\partial x_c}.
\end{equation}
This may be written
\begin{equation}
    \vv=
    \left|\begin{array}{ccc}
1 & 1 & 1 \\
\frac{\partial h_x}{\partial \alpha} & \frac{\partial h_y}{\partial \alpha} & \frac{\partial h_z}{\partial \alpha} \\
\frac{\partial h_x}{\partial \beta} & \frac{\partial h_y}{\partial \beta} & \frac{\partial h_z}{\partial \beta}
\end{array}\right|
\nabla{g_\alpha} \times \nabla{g_\beta}.
\end{equation}
The determinant is the area element on the unit sphere, so
\begin{equation}
    \vv=\left(\nabla g_\alpha \times \nabla g_\beta\right) D\left(g_\alpha, g_\beta\right).
\end{equation}

\section{Derivation of the isotrope field in arbitrary dimension}
\label{app:ndim}
For completeness, and as it may be of interest for broader applications of isotropes, we present a derivation of the expression for the isotrope field corresponding to a vector field in $d$-dimensional Euclidean space.

We take a set of coordinates $\{\phi_1,\dots\phi_{d-1}\}$ over $S^{d-1}$, and write a volume form over $S^{d-1}$ as
\begin{equation}
    \omega =D(\phi_{1},\dots \phi_{d-1})d\phi_{1}\wedge\dots\wedge d\phi_{d-1}.
\end{equation}
The standard choice with spherical coordinates would define $D$ as follows, although we do not require this definition.
\begin{equation}
    D(\phi_{1},\dots \phi_{d-1})=\sin ^{d-2}(\phi_{1})\sin ^{d-3}(\phi_{2})\dots \sin (\phi_{d-2}).
\end{equation}

As before, we are seeking some $\vv$ such that
\begin{equation}
    \vv=\star g^*\omega
\end{equation}
where $g(\xx):\xx\mapsto \left(g_{\phi_1}(\xx)\dots g_{\phi_{d-1}}(\xx)\right)$ is the map from $\mathbb{R}^d$ to $S^{d-1}$ defined by the unit vectors of some vector field $\VV$.
\begin{equation}
    g(\mathbf{x}) = \frac{\VV(\mathbf{x})}{|\VV(\mathbf{x})|}.
\end{equation}

Consider some hypersurface $Y\subset \mathbb{R}^d$, without loss of generality chosen to be $Y=\{\xx\in\mathbb{R}^d|x_d=0\}$. We have
\begin{equation}
    \int_{Y}\vv\cdot da= \int_{Y}g^*\omega=\int_{g_*Y}\omega
= \int_{g_{*}Y} d\phi_{1}\dots d\phi_{d-1}D(\phi_{1},\dots \phi_{d-1}).
\end{equation}

Pulling this integral back into $Y$ from $S^{d-1}$ and taking the corresponding coordinate transformation:
\begin{equation}
    \int_{Y}g^*\omega=\int _{Y}dx_{1}\dots dx_{d-1}\epsilon^{i_{1}\dots i_{d-1}}(\partial_{x_{1}}\gp{i_{1}})\dots(\partial_{x_{d-1}}\gp{i_{d-1}})D(\gp{1},\dots \gp{d-1}),
\end{equation}
from which we conclude:
\begin{equation}
    g^*\omega(\mathbf a_{1},\dots \mathbf a_{d-1})=  \epsilon^{i_{1}\dots i_{d-1}}(\partial_{\mathbf a_{1}}\gp{i_{1}})\dots(\partial_{\mathbf a_{d-1}}\gp{i_{d-1}})D(\gp{1},\dots \gp{d-1}).
\end{equation}

We can then compute the components of $\vv$. We start from the formula:
\begin{equation}
    \vv= \star g^*\omega. 
\end{equation}
Applying the definition of the Hodge dual and contracting $g^*\omega$ with vectors $\aaa_1,\dots\aaa_{d-1}$,
\begin{equation}
\begin{split}
\epsilon_{j_{1}\dots j_{d}}a_{1}^{j_{1}}\dots a_{d-1}^{j_{d-1}}v^{j_{d}}= & g^*\omega(\mathbf a_{1},\dots \mathbf a_{d-1}) \\
= & \epsilon^{i_{1}\dots i_{d-1}}(\partial_{\mathbf a_{1}}\gp{i_{1}})\dots(\partial_{\mathbf a_{d-1}}\gp{i_{d-1}})D(\gp{1},\dots \gp{d-1}) \\
\epsilon_{j_{1}\dots j_{d}}a_{1}^{j_{1}}\dots a_{d-1}^{j_{d-1}}v^{j_{d}}= & \epsilon^{i_{1}\dots i_{d-1}} \left( a_{1}^{j_{1}}\partial_{j_{1}}\gp{i_{1}} \right)\dots \left( a_{d-1}^{j_{d-1}}\partial_{j_{d-1}}\gp{i_{d-1}} \right) D(\gp{1},\dots \gp{d-1}).
\end{split}
\end{equation}
As $a_{1}^{j_{1}}\dots a_{d-1}^{j_{d-1}}$ is entirely general, we can remove it from both sides, yielding
\begin{equation}
\epsilon_{j_{1}\dots j_{d}}v^{j_{d}}= \epsilon^{i_{1}\dots i_{d-1}} \left( \partial_{j_{1}}\gp{i_{1}} \right)\dots \left( \partial_{j_{d-1}}\gp{i_{d-1}} \right) D(\gp{1},\dots \gp{d-1})
\end{equation}
We now multiply both sides by $\epsilon^{j_{1}\dots j_{d-1}k}$ and simplify
\begin{equation}
\begin{split}
\epsilon^{j_{1}\dots j_{d-1}k}\epsilon_{j_{1}\dots j_{d}}v^{j_{d}}= & \epsilon^{j_{1}\dots j_{d-1}k}\epsilon^{i_{1}\dots i_{d-1}} \left( \partial_{j_{1}}\gp{i_{1}} \right)\dots \left( \partial_{j_{d-1}}\gp{i_{d-1}} \right) D(\gp{1},\dots \gp{d-1}) \\
= & (d-1)!\ \delta^{k}_{j_{d}}v^{j_{d}},
\end{split}
\end{equation}
yielding the components of $\vv$:
\begin{equation}
v^k= \frac{1}{(d-1)!}\epsilon^{j_{1}\dots j_{d-1}k}\epsilon^{i_{1}\dots i_{d-1}} \left( \partial_{j_{1}}\gp{i_{1}} \right)\dots \left( \partial_{j_{d-1}}\gp{i_{d-1}} \right) D(\gp{1},\dots \gp{d-1}).
\end{equation}

This vector field can be proven to point along isotropes as follows.
\begin{equation}
\label{equ:ndimortho}
\begin{split}
\vv\cdot\nabla \gp{l}=v^k\partial_k\gp{l}=\frac{1}{(d-1)!}\epsilon^{j_{1}\dots j_{d-1}k}\epsilon^{i_{1}\dots i_{d-1}} \left( \partial_{j_{1}}\gp{i_{1}} \right)\dots \left( \partial_{j_{d-1}}\gp{i_{d-1}} \right) \left(\partial_k\gp{l}\right)D(\gp{1},\dots \gp{d-1}).
\end{split}
\end{equation}
For any nonzero term in this sum, there must be one element of the multi-index $\{i_1,\dots i_{d-1}\}$ that equals $l$. Without loss of generality, consider the case $i_{d-1}=l$.
\begin{equation}
\begin{split}
\epsilon^{j_{1}\dots j_{d-1}k}\left( \partial_{j_{d-1}}\gp{l} \right) \left(\partial_k\gp{l}\right)
=&\frac12\left(\epsilon^{j_{1}\dots j_{d-1}k}-\epsilon^{j_{1}\dots j_{d-2}kj_{d-1}}\right)\left( \partial_{j_{d-1}}\gp{l} \right) \left(\partial_k\gp{l}\right)\\
=&\frac12\epsilon^{j_{1}\dots j_{d-1}k}\left(\left( \partial_{j_{d-1}}\gp{l} \right) \left(\partial_k\gp{l}\right)- \left(\partial_k\gp{l}\right)\left( \partial_{j_{d-1}}\gp{l} \right)\right)\\
=& 0.
\end{split}
\end{equation}
Therefore, all terms in equation~\eqref{equ:ndimortho} must vanish, and $\vv$ is orthogonal to the gradients of all components of $g$. The coordinates are constant along stream lines of $\vv$, i.e. $\vv$ defines the isotrope field.

\section{Straightening External fields}
\label{app:straightening}
Our goal is to demonstrate a means of constructing a coordinate transformation which preserves the topology of vector fields and which yields a basis in which an arbitrary magnetic field is constant.
This method is applicable so long as all field lines start and end on the boundary of the domain under consideration.
We begin by noting that a vector field has constant coefficients in a particular coordinate system precisely when it commutes (with respect to the Lie bracket) with the basis vector fields.
We can generate a vector field $\XX$ that commutes with a given magnetic field $\BB$ by choosing its values on some surface and Lie dragging it along the magnetic fieldlines passing through the surface.
This amounts to integrating the differential equation
\begin{equation}\label{liedrag}
B_i\left(\partial_iX_j\right)=\left(\partial_iB_j\right)X_i. 
\end{equation}
for a particular choice of Cauchy data, which by definition gives a solution such that $[\BB,\XX]_j=B_i\left(\partial_iX_j\right)-\left(\partial_iB_j\right)X_i=0$.
By choosing two commuting vector fields $\XX$, $\YY$ lying on the boundary such that $\BB\cdot \XX \times \YY=1$ and Lie dragging as above, we construct a coordinate basis $\left\{\XX,\YY,\BB\right\}$ in which $\BB=(0,0,1)$ everywhere. Note that their mutual commutation is required for the vector fields to generate consistent coordinates.
\\
This process can also be considered from the perspective of matrix transformations on tangent spaces of $\mathbb{R}^3$.
Equation \ref{liedrag} can be written
\begin{equation}\frac{d}{ds}\XX=\MM\cdot \XX.
\end{equation}
  where $s$ is a fieldline parameter such that $\frac{dx_i}{ds}=B_i$ and $\mathsf{M}_{ij}=\partial_jB_i$ is the Jacobian of the magnetic field.
  We are seeking some transformation matrix $\TT(x)\in SL(3,\mathbb{R})$ that maps an initial vector on the boundary to its Lie dragging to some point $\xx$: $\XX(\xx)=\TT(\xx)\cdot \XX_0$.
  This yields
  \begin{equation}\frac{d}{ds}\TT=\MM\cdot \TT,
  \end{equation}
  with the solution
  \begin{equation}\TT=\mathcal{T}\left\{e^{\int \MM ds}\right\}
    =\sum_{n=0}^\infty\int_0^L\int_0^{s_n}\int_0^{s_{n-1}}...\int_0^{s_2}
  \MM(s_n)...\MM(s_1)ds_1...ds_n
  \end{equation}
  where $\mathcal{T}$ denotes the time-ordering of the integral along the path.
  Noting that $\int \mathsf{M}_{ij} ds=\int \mathsf{M}_{ij} \BB\cdot dl$, we see that we are integrating a Lie algebra valued one form $\mathsf{M}_{ij} B_{\alpha}:T\mathbb{R}^3\rightarrow \mathfrak{sl}(3,\mathbb{R})$ over a curve that happens to be the streamline of the magnetic field.
  The resulting matrix in $SL(3,\mathbb{R})$ defines a basis transformation at the end point (and at any intermediate point along the path) that maps $\BB$ at the start point to $\BB$ at the endpoint and functions to Lie drag vectors. This means of transporting vector fields along field lines might be extended to arbitrary paths, producing a gauge connection on the tangent bundle.

\section{Jacobian of the dipole field}
\label{app:dipole}
Here we briefly show for completeness that the Jacobian of a dipole field with moment pointed along $\hat{z}$ is negative for $z>0$ and positive for $z<0$.
Written in Cartesian coordinates, the unit dipole field is
\begin{equation}
    \BB=
    \left(\frac{3 x z}{\left(x^2+y^2+z^2\right)^{5/2}},\frac{3 y z}{\left(x^2+y^2+z^2\right)^{5/2}},\frac{2 z^2-x^2-y^2}{\left(x^2+y^2+z^2\right)^{5/2}}\right)
\end{equation}
with Jacobian
\begin{equation}
    \MM=
\frac1{(x^2+y^2+z^2)^{7/2}}
\left(
\begin{array}{ccc}
 3 z \left(-4 x^2+y^2+z^2\right) & -15 x y z & 3 x \left(x^2+y^2-4 z^2\right) \\
 -15 x y z & 3 z \left(x^2-4 y^2+z^2\right) & 3 y \left(x^2+y^2-4 z^2\right) \\
 3 x \left(x^2+y^2-4 z^2\right) & 3 y \left(x^2+y^2-4 z^2\right) & 9 x^2 z+9 y^2 z-6 z^3 \\
\end{array}
\right).
\end{equation}
The determinant is
\begin{equation}\label{eq:dipolejac}
    |\MM|=
    -\frac{27 z \left(x^2+y^2+2 z^2\right)}{\left(x^2+y^2+z^2\right)^{15/2}}.
\end{equation}
From equation~\eqref{eq:dipolejac} it is clear that ${\rm sign}\left(|\MM|\right)=-{\rm sign}(z)$

\end{document}